\documentclass[11pt] {article}
\usepackage{latexsym,amssymb,epsf,rotating,axodraw}
\newcommand{\newc}{\newcommand}
\newcommand{\lsim}{\, \, \raisebox{-0.8ex}{$\stackrel{\textstyle <}{\sim}$ }}
\newcommand{\gsim}{\, \, \raisebox{-0.8ex}{$\stackrel{\textstyle >}{\sim}$ }}
\newc{\ra}{\rightarrow}
\newc{\lra}{\leftrightarrow}
\newc{\lam}{\lambda}
\newc{\ie}{i.e.}
\newc{\lsp}{{{\tilde\chi}}}
\newc{\rttwo}{\sqrt{2}}
\newc{\emax}{E_\lsp^{\rm max}}
\newc{\half}{\frac{1}{2}}
\newc{\cw}{\cos \theta_w}
\newc{\sw}{\sin \theta_w}
\newc{\gev}{\,{\rm GeV}}
\newc{\mev}{\,{\rm MeV}}
\newc{\phhigh}{\phantom{\huge I}}

\begin{document}

\title{Supernov{\ae} and Light Neutralinos: SN1987A 
Bounds on Supersymmetry Revisited}

\author{H.~K.~Dreiner$^a$, C.~Hanhart$^b$, U.~Langenfeld$^a$, and
D.~R.~Phillips$^c$}
\date{\small{$^a$Physikalisches Institut, Universit\"at Bonn, Nu\ss allee 12,
53113 Bonn, Germany;} \\
  \small{$^b$Institut f\"ur Kernphysik, Forschungszentrum J\"ulich, 52428
J\"ulich, Germany;} \\
  \small{$^c$Department of Physics and Astronomy, Ohio University,
    Athens, OH 45701, USA;}}
\maketitle

\vspace{-7cm}
\hspace{11.5cm}{\parbox[t]{5cm}{BONN-TH-2003-01 \\ FZJ-IKP-TH-2003-3}}
\vspace{+7cm}

\begin{abstract}
\noindent
For non-universal gaugino masses, collider experiments do not provide
any lower bound on the mass of the lightest neutralino.  We review the
supersymmetric parameter space which leads to light neutralinos,
$M_\lsp \lsim {\cal O}(1\gev)$, and find that such neutralinos are
almost pure bino.  In light of this, we examine the neutralino lower
mass bound obtained from supernova 1987A (SN1987A).  We consider the
production of binos in both electron-positron annihilation and
nucleon-nucleon binostrahlung.  For electron-positron annihilation, we
take into account the radial and temporal dependence of the
temperature and degeneracy of the supernova core.  We also separately
consider the Raffelt criterion and show that the two lead to
consistent results.  For the case of bino production in $NN$
collisions, we use the Raffelt criterion and incorporate recent
advances in the understanding of the strong-interaction part of the
calculation in order to estimate the impact of bino radiation on the
SN1987A neutrino signal. Considering these two bino production
channels allows us to determine separate and combined limits on the
neutralino mass as a function of the selectron and squark masses.  For
$M_\lsp \sim 100 \mev$ values of the selectron mass between 300 and
900 GeV are inconsistent with the supernova neutrino signal. On the
other hand, in contrast to previous works, we find that SN1987A
provides almost no bound on the squark masses: only a small window of
values around 300 GeV can be excluded, and even then this window
closes once $M_\lsp \gsim 20 \mev$.

\end{abstract}

\section{Introduction}
If the supersymmetry-breaking, electroweak, gaugino masses fulfill
the grand-unified mass relation:
\begin{equation}
M_1=\frac{5}{3}\tan^2\!\theta_w M_2,
\label{assump}
\end{equation}
then, because LEP chargino and neutralino-pair-production
searches set a lower bound on $M_2$, there is a concomitant lower bound on
$M_1$. These constraints on the mass matrix result (for the case of
conserved R-parity) in a limit on the mass of the lightest neutralino,
$\lsp$~\cite{Abdallah}:
\begin{equation}
M_{\lsp}\geq 46\gev.
\label{bound}
\end{equation}

However, Eq.~(\ref{assump}) is not an inescapable consequence of
unification. For example, unification might occur through a string
theory without a simple gauge group~\cite{Polchinski:rq}.
Refs.~\cite{Choudhury:1999tn,Dedes:2001zi} showed that if the
assumption (\ref{assump}) is dropped and $M_1$ and $M_2$ are
considered as independent free parameters then there is {\it currently
no lower experimental bound on the neutralino mass from collider
experiments} ~\footnote{The sensitivity at LEP2 to the cross section
$\sigma(ee\ra\lsp^0_1\lsp^0 _2)$ in the case where $M_{\lsp^0_1}$ and
$M_{\lsp^0_2}$ are free parameters \cite{Abdallah} can be compared
with the required sensitivity shown in Fig. 6 of
Ref.~\cite{Choudhury:1999tn}.  It is then clear that the light
neutralino is not excluded by LEP2 but could possibly be produced at a
next-generation collider.}.  In light of this situation, here we
reconsider the bounds on neutralino properties which can be determined
from supernova 1987A
(SN1987A)~\cite{Grifols:fw,Ellis:1988aa,Lau:vf,Kachelriess:2000dz}.

The basic idea is that in a supernova, neutralinos with masses of order
the supernova core temperature, $T_c={\cal O}(30\mev)$, can be
produced in large numbers via electron-positron annihilation
\cite{Grifols:fw,Lau:vf} and nucleon-nucleon ($NN$) ``neutralinostrahlung''
\cite{Ellis:1988aa}:
\begin{eqnarray}
e^++e^-&\longrightarrow& \lsp+\lsp\,, \label{Xs-e}\\
N+N&\longrightarrow& N+N+\lsp+\lsp\,. \label{Xs-n}
\end{eqnarray}
Once produced, the neutralinos have a mean-free-path, $\lam_\lsp$,
in the supernova core which is determined via the cross sections
for the processes~\cite{Grifols:fw,Ellis:1988aa,Lau:vf}:
\begin{eqnarray}
\lsp +e &\longrightarrow& \lsp+e\,,\label{scatter1} \\
\lsp + N &\longrightarrow& \lsp+N\,,
\label{scatter2}
\end{eqnarray}
as well as the electron and nucleon densities.  If $\lam_\lsp$ is of
order the core size, $R_c={\cal O}(10\,{\rm km})$, or larger, the
neutralinos escape freely and thus cool the supernova rapidly.  As the
supernova temperature drops, the neutrino scattering cross section
also drops and the neutrinos are no longer trapped, thereby further
hastening the cooling of the supernova.  Thus neutralino cooling could
significantly shorten the supernova neutrino signal~\cite{Gandhi:bq},
in disagreement with the observation from SN1987A by the Kamiokande
and IMB collaborations~\cite{Hirata:1987hu,Bionta:1987qt}.  However,
if the neutralinos have masses $M_\lsp$ much greater than the
supernova core temperature $T_c$ then their production is
Boltzmann-factor suppressed and they affect the cooling negligibly,
independent of $\lam_\lsp$.  Demanding that $M_\lsp$ be large enough
that neutralino-cooling does not markedly alter the neutrino
signal---particularly its time-structure---allows us to set a lower
limit on the neutralino mass.  As we discuss in detail below, this
limit depends on the squark and selectron masses which determine the
relevant neutralino cross sections for the production processes
(\ref{Xs-e}) and (\ref{Xs-n}), as well as those for the scattering processes
(\ref{scatter1}) and (\ref{scatter2}).

A full treatment of this physics requires the implementation of the
neutralinos into the supernova simulation code.  This is well beyond
the scope of this paper.  Instead we invoke two different analytic
criteria to estimate whether neutralino emission will affect the
detected neutrino signal significantly.  The first test we consider is
to require the integrated supernova energy emitted through the
neutralino channel to be less than
\begin{equation}
\emax =10^{52}\,{\rm erg}\,,
\label{energy}
\end{equation}
\ie\ much smaller than the energy emitted by all neutrino species:
$E_\nu \approx 3.0 \times 10^{53}\,{\rm erg}$ \cite{Raff}.  The exact
number we choose here is somewhat arbitrary--not least since it is
subject to our ignorance of the total energy released in SN1987A. We
will discuss the dependence of our neutralino mass bounds on $\emax$
below.

Detailed supernova simulations with axions suggest that another way to
place bounds on these energy-loss mechanisms is the ``Raffelt
criterion''~\cite{Raff}. This states that ``exotic'' cooling
processes, such as those considered here, will not alter the neutrino
signal observably, provided that their emissivity, $\dot{\cal E}$,
obeys:
\begin{equation}
\dot{\cal E} < 10^{19}~{\rm ergs/g/s}.
\label{raffelt}
\end{equation}
If their emissivity is larger than this they will remove sufficient
energy from the explosion to invalidate the current understanding of a
Type-II supernova neutrino signal. With this simple criterion the
results of the detailed simulations in
Refs.~\cite{Burrows:1988ah,HPPR}---where the exotic cooling mechanisms
considered were axions~\cite{Burrows:1988ah} and KK-graviton
emission~\cite{HPPR}---can be reproduced.  These detailed simulations
also show that the feature of the SN1987A neutrino pulse which is most
affected by the presence of alternative cooling mechanisms is the
temporal distribution. Losing $10^{52}$ ergs to neutralinos during the
first $\sim$ 1 second after bounce, as in our first criterion
(\ref{energy}), might not greatly affect neutrino temporal and spectral
features. But, at later times, the loss of just a fraction of this
energy could have a dramatic impact---by ending the period of
diffusive neutrino cooling much earlier than would otherwise be the
case.  In this paper, we examine the impact of the process
(\ref{Xs-e}) using first Eq.~(\ref{energy}) and then the Raffelt
criterion, Eq.~(\ref{raffelt}), and compare the results. We find good
agreement between the two.

\medskip

Neither of these criteria are particularly suitable in the case that
neutralinos have a small mean-free-path, $\lam_\lsp\lsim {\cal
O}(R_c)$, since the neutralinos' contribution to the proto-neutron
star cooling is then diffusive, rather than radiative.  The treatment
of this ``trapped'' regime is beyond the scope of this study. In
Sec.~\ref{trap} we make an estimate of the conditions for trapping
and then refer the reader to the literature for bounds on SUSY
parameters under these
conditions~\cite{Grifols:fw,Ellis:1988aa,Lau:vf}.  In this work, we
focus on estimating the emitted energy in neutralinos, assuming they
are weakly coupled to matter, and then applying the criteria
(\ref{energy}) and (\ref{raffelt}).

\medskip

Previous authors~\cite{Grifols:fw,Ellis:1988aa,Lau:vf} considered the
case of a stable neutralino and were thus restricted to the mass
regions~\cite{Ellis:1983ew}%
~\footnote{The lower bound has recently been reconsidered
\cite{Belanger:2002nr,Hooper:2002nq,Bottino:2002ry} in the light of
new data from LEP and $(g-2)_\mu$. Associating the neutralino with the
dark matter of our universe these works obtain $M_\lsp>{\cal
O}(5\gev)$. The upper bound in Eq.~(\ref{eq:cosmobd}) was not
considered in
Refs.~\cite{Belanger:2002nr,Hooper:2002nq,Bottino:2002ry}, although in
fact sufficiently small neutralino masses are always allowed by the
closure constraint, independent of the sfermion masses.  However this
light a neutralino would constitute a sizable {\it hot} dark matter
component, whose presence structure-formation arguments strongly
disfavour~\cite{Davis:rj}.}:

\begin{equation}
M_\lsp\lsim 100~{\rm eV}\,\qquad {\rm or} \qquad M_\lsp \gsim
500\mev\,,
\label{eq:cosmobd}
\end{equation}
in order to avoid an over-closed universe. The heavier neutralinos,
$M_\lsp>500\mev$, are irrelevant for supernov{\ae} and thus
Refs.~\cite{Grifols:fw,Ellis:1988aa,Lau:vf} focused on a massless
neutralino. In this work we consider the range:
\begin{equation}
0 \leq M_\lsp \leq 200\mev\,.
\end{equation}
We avoid over-closure of the universe by allowing for the possibility
of a small amount of R-parity violation
\cite{Dreiner:pe,Dreiner:1997uz}, and thus assume that the
neutralinos, although stable on the time scale of the supernova and
collider experiments, are not stable on cosmological timescales. For
such neutralinos, which we call ``quasi-stable'', the mass restriction
(\ref{eq:cosmobd}) does not apply~\footnote{A detailed analysis of
  cosmologically allowed neutralino lifetimes
  \cite{Dreiner:1997uz,Ellis:1990nb,cosmo} will be given elsewhere.}.

\medskip

To summarize: in this work we consider a general lower mass bound on
quasi-stable neutralinos as a function of the squark and selectron
masses. We assume the bounds from the relic density are avoided
through small R-parity violating couplings.  This is similar in
outlook to Ref.~\cite{Kachelriess:2000dz} where, in light of the
Karmen time anomaly~\cite{Armbruster:nr}, the impact on SN1987A of a
neutralino with the specific mass ($M_\lsp=34\mev$) was examined.  We
extend this work to a general neutralino.  We also differ from
previous work in the following more technical points
\begin{enumerate}
\item Throughout we consider a bino neutralino instead of a photino,
  since in Refs.~\cite{Choudhury:1999tn,Dedes:2001zi} it was shown that
  a light neutralino, $M_\lsp\lsim 5\gev$, must be dominantly bino in
  order to be consistent with LEP results. This leads to a
  substantially weaker effective coupling to nucleons and thus to
  weaker bounds on the supersymmetric-particle masses.

\item For the case of electron-positron annihilation and the criterion
  Eq~(\ref{energy}), we consider the radial and temporal dependence of
  the temperature and the degeneracy in the core
  \cite{Burrows:me}. This gives a more realistic estimate of this
  contribution.  It also gives a {\it stricter} bound than some
  previous work \cite{Ellis:1988aa} since the temperature in the
  outermost region of the star ($r\approx\frac{9}{10}R_c$), where most
  of the neutralinos coming from annihilation are produced, is
  somewhat higher than the temperature considered by, for instance,
  Ref.~\cite{Lau:vf}.  However, it gives a weaker bound than in the
  earlier work by Ellis~{\it et al.}, Ref.~\cite{Ellis:1988aa}, where
  a significantly higher core temperature was chosen. Since all
  emissivities vary rapidly with temperature this results in a marked
  difference in the bounds on the supersymmetric-particle masses.

\item The calculation of the $NN \rightarrow NN \lsp\lsp $ emissivity
by Ellis~{\it et al.} was based on the computation of the rate for the
related neutrinostrahlung process: $NN \rightarrow NN \nu {\bar
\nu}$~\cite{Ellis:1988aa}. For this Ellis~{\it et al.} used the
pioneering calculation of Friman and Maxwell, Ref.~\cite{FM79}. A
recent model-independent treatment of $NN$ dynamics in the production
of axial radiation---e.g. by reactions such as $NN \rightarrow NN \nu
\bar{\nu}$ or $NN \rightarrow NN \lsp\lsp$---suggests that
Ref.~\cite{FM79} overestimates supernova emissivities by about a
factor of four~\cite{HPR,Ti02,vD03}. This loosens the bound previously
obtained from SN1987A on the relevant supersymmetric-particle masses.
  
\item Since previous authors only considered very light neutralinos,
  $M_\lsp$ was neglected in computing the total neutralino emissivity.
  We calculate the full $M_\lsp$-dependence of $\dot{\cal E}_\lsp$. In
  particular, for $M_\lsp \gsim T_c$ the mass dependence is very
  strong. Indeed, ultimately it is exactly this strong mass dependence which
  we use to derive a bound on $M_\lsp$.

\item We also combine the emissivities from both electron-positron
  annihilation and nucleon-nucleon neutralinostrahlung to get a bound
  on supersymmetric-particle masses from both sources of neutralino
  radiation.

\end{enumerate}

The outline of the paper is as follows. In Sec.~\ref{neutralino}, we
discuss the possibility of a light neutralino in the MSSM.  In
Sec.~\ref{ep}, we determine the bounds which can be found from
electron-positron annihilation. This allows us to exclude certain
regions of the joint parameter space of selectron and neutralino
masses. In Sec.~\ref{nn-eff}, we determine the effective
neutralino-nucleon coupling. In Sec.~\ref{nn}, we use this to compute
the general bound obtained from $NN \rightarrow NN \lsp\lsp$ as a
function of $M_\lsp$ and $M_{\tilde q}$. This analysis can be applied
to the production of any particle pair which is axially coupled to
nucleons. In Sec.~\ref{trap} we discuss the effect that neutralino
trapping due to matter and/or gravitational interactions has on these
bounds. Finally, in Sec.~\ref{conc} we combine the results for $e^+
e^-\rightarrow\lsp\lsp$ and $NN \rightarrow NN\lsp\lsp$ in order to
get overall information on the regions of MSSM parameter-space which
are excluded, and offer our conclusions.

\section{A Light Neutralino in the MSSM}
\label{neutralino}

Before we discuss in detail the bounds we obtain from SN1987A on a
light neutralino, we consider whether such a neutralino is consistent
within the MSSM and with existing collider data. In
Ref.~\cite{Choudhury:1999tn} it was shown that there are no laboratory
bounds on a neutralino with mass $M_\lsp=34\mev$, provided it is
dominantly bino---mainly because a bino neutralino does not couple to
the $Z^0$-boson at tree-level.  For a neutralino with mass
$M_\lsp<{\cal O}(200\mev)$---which we consider here---none of the
bounds in Refs.~\cite{Choudhury:1999tn,Dedes:2001zi} depend
sensitively on the mass. We thus expect such a neutralino to be
consistent with all terrestrial data~\footnote{A detailed analysis can
be found in Ref.~\cite{slavich}.}.

\begin{figure}[t]
\vspace*{11cm}
\includegraphics{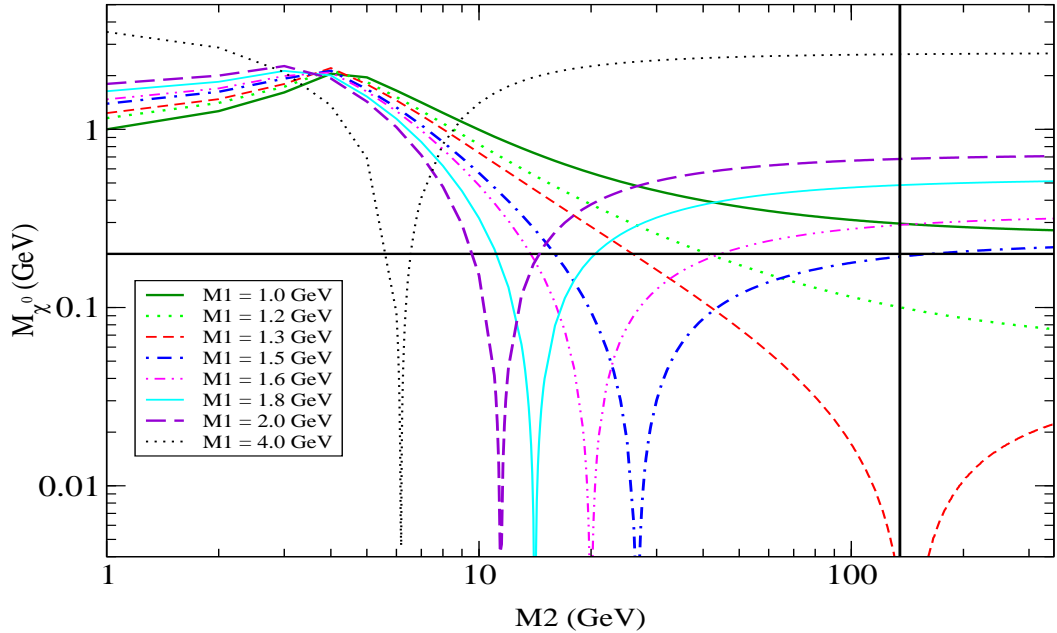}
\vspace*{-2.5cm}
\caption{{\it The neutralino mass as a function of $M_2$ for various
    values of $M_1$ and with $\tan\beta=10$, $\mu=300\gev$. The
    downward spikes correspond to true zeros of the neutralino mass.
    The bold horizontal line indicates the value $M_\lsp= 200\mev$,
    below which we consider in this paper.  To the right of the bold
    vertical line at about $M_2=120\gev$ the chargino mass satisfies
    the lower mass bound from LEP2.}}
\label{parameter}
\vspace{-0.4cm}
\end{figure}

We have gone beyond Refs.~\cite{Choudhury:1999tn,Dedes:2001zi}, and
looked in more detail at where in the MSSM parameter space a
neutralino with $M_\lsp \leq 200$ MeV can be obtained. We have found
new parameter regions which are shown in Fig.~\ref{parameter}. Note
particularly that for a fixed value of $M_1$ there can be more than
one value of $M_2$ leading to a specific $M_\lsp$. It is also clear
that, as was pointed out in Ref.~\cite{lykken}, neutralino zero modes
are possible.  These are the downward spikes in Fig.~\ref{parameter}.
There are in fact extensive regions in the $M_1-M_2$ parameter space
where a light neutralino can be obtained.  However, if we take into
account the lower mass bound on the chargino from LEP2 ($M_2>120\gev$
), we see that for\footnote{$\tan\beta=v_1/v_2$ is the ratio of the
vaccum expectation values of the two neutral CP-even Higgs bosons in
the minimal supersymmetric standard model. $\mu$ is the mixing
parameter of the Higgs fields in the superpotential.}  $\tan\beta=10$
and $\mu=300\gev$ a light neutralino is only obtained for the specific
range $1\gev <M_1<1.5\gev$.  (These parameter values will be shifted
by the one-loop radiative corrections to the gaugino masses.) This
involves a relative fine-tuning of $10^{-2}$ between $M_1$ and $M_2$.
While this is perhaps {\ae}sthetically displeasing, it is by no means
forbidden.

In this parameter range, the neutralino is more than 98\% bino.  In
the following we shall thus work with a pure bino. Our only free
supersymmetric parameters are then $M_\lsp$ and the selectron
and squark masses.

\section{Electron-Positron Annihilation to Neutralinos}
\label{ep}
We first discuss the case of free-streaming neutralinos.  We defer the
issue of trapped neutralinos to Sec.~\ref{trap}, where we consider
their scattering off electrons and nucleons in the supernova
environment, as well as gravitational trapping.

\subsection{Emissivity for Free Streaming Neutralinos}

The neutralinos are produced via electron-positron annihilation
\begin{equation}
e^+(p_1)+e^-(p_2)\ra \lsp(k_1)+\lsp(k_2)\,,
\label{Xs-e2}
\end{equation}
where we have indicated the four-momenta of the particles.  The
dominant process proceeds via $t$- and $u$-channel selectron exchange.
The energy thus emitted by the supernova per unit time and unit volume
is the emissivity
\begin{eqnarray}
\dot{\cal E}(M_\lsp,T_c,\eta)\equiv
\frac{d {\cal E}}{dt}=\int \frac{d^3p_1 d^3p_2}{(2 \pi)^6}
\,f_1f_2\,(E_1+E_2)\; |\Delta{\bf v}| \, \sigma(e^++e^-\ra
\lsp+\lsp) \,.
\label{emiss-ep}
\end{eqnarray}
Here $E_1+E_2=E_3+E_4$ is the combined energy of the emitted
neutralinos.
The Fermi-Dirac distributions are given by
\begin{equation}
f_i=\frac{1}{e^{(E_i\pm\mu_i)/T_c}+1}\;,
\label{eq:FDdist}
\end{equation}
where $\mu_i$ is the chemical potential and $T_c$ the temperature in
the supernova.  In the following, we shall write $\eta\equiv\mu/T_c$
for the degeneracy of the electrons.  We have neglected the Pauli
blocking of the final state neutralinos: $(1-f_3)(1-f_4)$. In
Eq.~(\ref{emiss-ep}), $\sigma(e^++e ^-\ra \lsp+\lsp)$ is the
(free-space) cross section for the process (\ref{Xs-e2}) and
$|\Delta{\bf v}|$ is the absolute value of the relative M{\o}ller
velocity
\begin{eqnarray}
{v}_{\rm{M\!{\not \,o}l}} = \sqrt{(\mathbf{v}_1 - \mathbf{v}_2)^2 -
            (\mathbf{v}_1 \times \mathbf{v}_2)^2} \;
\stackrel{{v_i\ra1}}{\longrightarrow}\;(1-\cos\theta)\;.
\end{eqnarray}
(${\bf v}_i$ are the velocities of the incoming electron and positron
and $\theta$ is the angle between them.) In the case where the
selectron masses are degenerate, $M_{{\tilde e}_L}=M_{{\tilde
e}_R}=M_{\tilde e}$ the cross section is given by
\begin{equation}
\sigma(e^++e^-\ra \lsp+\lsp)=\frac{17\pi\alpha^2 s}
{24 \cos^4\theta_w
M_{\tilde e}^4} \left(1-\frac{4M_\lsp^2}{s}\right)^{3/2}\!\!\!,
\label{Xsec}
\end{equation}
where we have used $m_e^2 \ll s \ll M_{\tilde e}^2$. Here $\alpha$ is
the fine structure constant, $\theta_w$ is the electroweak mixing
angle and $s$ is the center-of-mass energy squared. Replacing the
bino-coupling by the photino coupling this agrees with
Ref.~\cite{haber}.

The emissivity in Eq.~(\ref{emiss-ep}) is a function of the neutralino
mass via the cross section (\ref{Xsec}). It is a function of the
supernova temperature and $\eta$ via the Fermi-Dirac distributions
(\ref{eq:FDdist}).

\subsection{Total Emitted Energy}

In order to determine the total energy emitted in the neutralino
channel we must integrate the emissivity over the time, $t_0$, during
which neutralinos are emitted and over the volume of the supernova
core
\begin{equation}
E_\lsp(M_\lsp)=\int_0^{t_0} \!dt\int \!d^3r \;\dot{\cal E}
(M_\lsp,T_c({\bf r},t),
\eta({\bf r},t)) \equiv \int_0^{t_0} \!dt \; P(M_\lsp,t)\,.
\label{eq:Edef}
\end{equation}
When performing these integrations care must be taken since the
temperature and the degeneracy depend on the radius and on the time.
We use the temperature distribution given in Fig.~1 of
Ref.~\cite{Burrows:me}.  There the temperature is shown as a function
of the enclosed baryon mass.  In Fig.~5 of Ref.~\cite{Burrows:me} we see
that for $t \geq 250$ ms the density is constant.  This makes the
conversion of $d^3r$ to $dM$ straightforward. Here we adopt a core
radius of $R_c=13$ km and a mass of $M_{\rm SN}=1.4\,M_{\odot}$ and obtain
\begin{equation}
\rho\approx 3 \times 10^{14}~{\rm g}/{\rm cm^3}.
\end{equation}
(Throughout we assume radial symmetry~\cite{Burrows:me}.)

The time $t=0$ of Ref.~\cite{Burrows:me} corresponds to the time when
the incoming shock wave stops and bounces outwards again.  The
boundary conditions at this time are not well known \cite{others}.
However, the exact shape of the initial conditions has little effect
on the subsequent evolution and after 0.5~s the distributions are
reliable.  For the first second we have used the $t=0.5\,$s
distributions of Ref.~\cite{Burrows:me}.  For longer time scales we
have used the subsequent radial distributions for $T_c({\bf r},t)$ and
$\eta({\bf r},t)$. However, as we will show below, most of the energy
is emitted during the first second after the bounce and so ultimately
we will use $t_0=1$ second in (\ref{eq:Edef}) to derive our neutralino
mass bounds. This means that we are demanding that the prompt
neutralino pulse in the first second does not have a total energy
greater than the bound (\ref{energy})~\footnote
{We have compared our integration with the ratio presented in
Ref.~\cite{Kachelriess:2000dz} and agree \cite{privat}. We thank
Michael Kachelriess for discussions on this point.}.
We discuss below how the neutralino-mass bound depends on the value chosen for
 $t_0$.

\subsection{Results}

\begin{figure}[t]
\vspace{8.cm}
\includegraphics{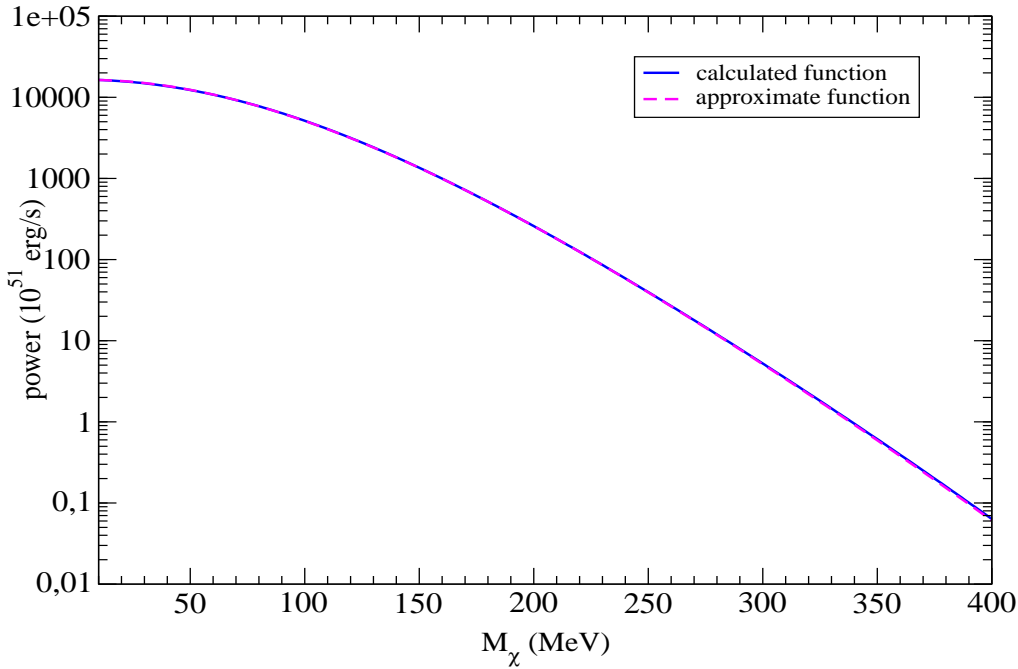}
\caption{{\it Neutralino power $P_\lsp(M_\lsp,t)$ for $t=0.5$ seconds as a
function of the neutralino mass, $M_\lsp$, with the selectron mass
$M_{{\tilde e}} =200\gev$. The solid curve shows our numerical
computation while the dashed curve represents the fit
Eq.~(\ref{fit-e}).  The two curves are almost indistinguishable.  }}
\label{power}
\end{figure}

Before presenting the total emitted energy, we consider the power
radiated in the neutralino channel.  In Fig.~\ref{power}, we show as
the solid curve the power of the neutralino emission at $t=0.5\;$s as
a function of the neutralino mass, $P(M_\lsp,0.5~{\rm s})$. We have
fixed the selectron mass to $M_{{\tilde e}}=200\gev$.  According to
Eq.~(\ref{Xsec}), the power scales as $M_{{\tilde e}}^{-4}$.  A good
fit to the solid curve throughout most of the neutralino mass range is
given by
\begin{eqnarray}
P(x) = C \left(\frac{200\gev}{M_{\tilde e}}\right)^4
\exp(a_4 x^4 + a_3 x^3 + a_2 x^2 + a_1 x)\,,\qquad x\equiv \frac{M_\lsp}{
{\rm MeV}}\,,
\label{fit-e}
\end{eqnarray}
where the constants are given by
\begin{eqnarray}
a_1&=& 1.71 \times 10^{-3} \,,\qquad \;\,
a_2\,=\, -1.58 \times 10^{-4}  \nonumber\\
a_3&=& 2.695 \times 10^{-7}\,,\qquad 
a_4\,=-1.99 \times 10^{-10}\, 
\,,\qquad C= 1.6318 \times 10^{55}\;{\rm erg/s}.
\end{eqnarray}
The numerical fit in Eq.~(\ref{fit-e}) is shown as the dashed curve in
Fig.~\ref{power}. Throughout the mass range considered it agrees to
better than 1\% and the two curves are almost indistinguishable.

When using the Raffelt criterion it is also convenient to have a
parameterization for the emissivity
\begin{eqnarray}
{\dot{\cal E}}(M_\lsp) = D \left(\frac{200\gev}{M_{\tilde e}}\right)^4
\exp(b_4 x^4 + b_3 x^3 + b_2 x^2 + b_1 x)\,,\qquad x\equiv \frac{M_\lsp}{
{\rm MeV}}\,,
\end{eqnarray}
and the constants are given by
\begin{eqnarray}
b_1&=& 4.75 \times 10^{-4} \,,\qquad 
b_2\,=\, -2.25\times 10^{-4} \nonumber \\
b_3&=& 4.66 \times 10^{-7}\,,\qquad 
b_4\,=\, -4.02 \times 10^{-10}\,,\qquad  D= 9.0125 \times 10^{21}\;{\rm erg/g/s}.
\end{eqnarray}

\begin{figure}[t]
\vspace{8.cm}
\includegraphics{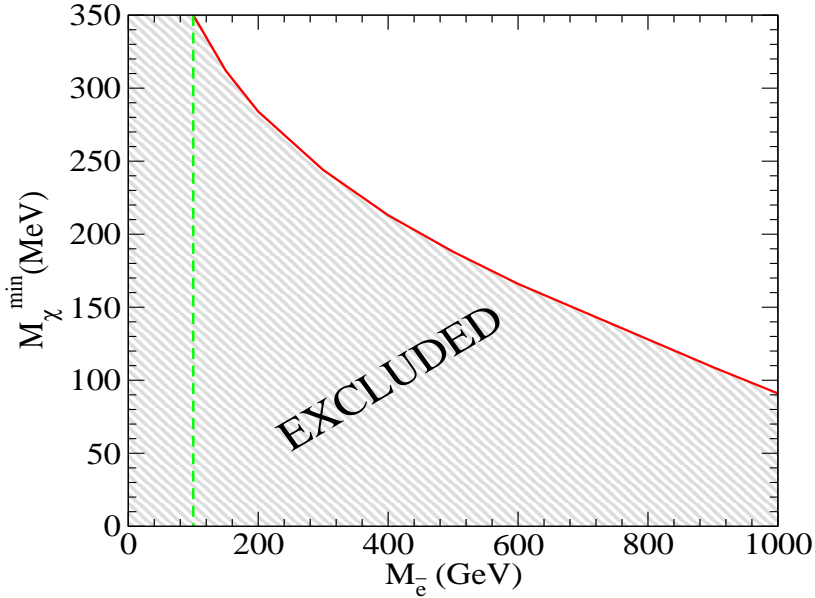}
\vspace{-0.3cm}
\caption{\it The dependence of the neutralino lower mass bound on the
  selectron mass.  Here we chose $\emax=10^{52}\,$erg and $t_0=1\,$s.
  The green-dashed line indicates the lower bound on the selectron
  mass from LEP2. Beyond $M_{\tilde e}=1$ TeV the exclusion curve
  drops rapidly. }
\label{selectron}
\vspace{-0.5cm}
\end{figure}

\begin{figure}[t]
\vspace{8.cm}
\includegraphics{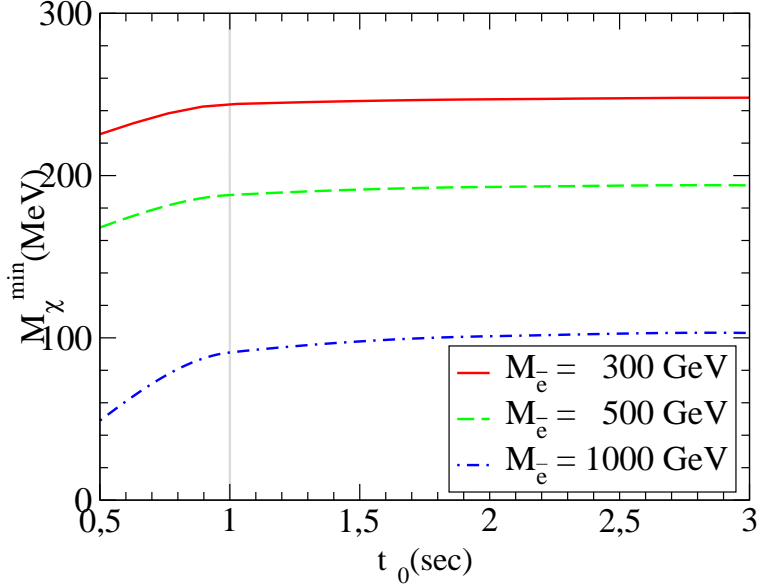}
\caption{{\it The minimum allowed value of $M_\lsp$, $M_\lsp^{\rm min}$,
    as a function of the time over which we integrate, $t_0$. The
    curves correspond to three different values of the selectron mass:
    $M_{\tilde e}=300,\, 500,\,1000\gev$.  Here
    $\emax=10^{52}\,$erg and the neutralino is pure bino. Our choice
    of $t_0=1$ second is indicated by the grey vertical line.}}
\label{time-dep}
\end{figure}

\begin{figure}[t]
\vspace{8.cm}
\includegraphics{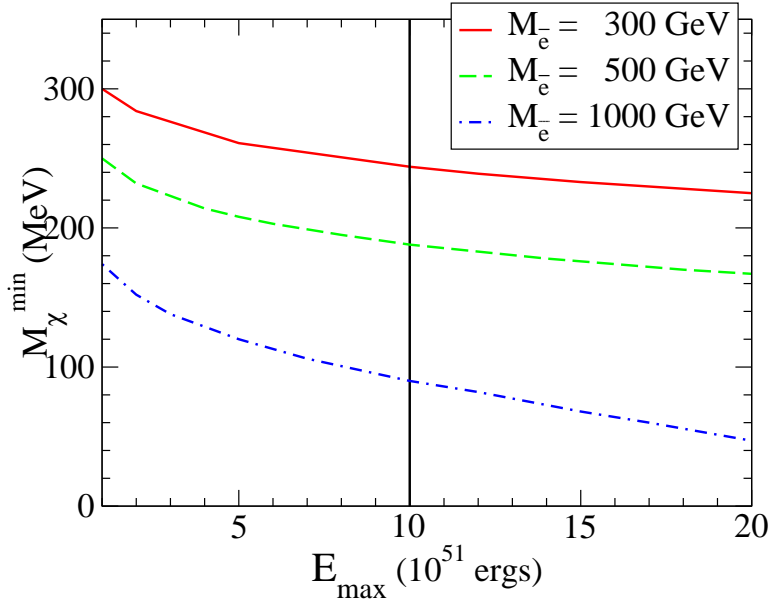}
\caption{{\it The neutralino
    mass bound as a function of $\emax$ for three different values of
    the selectron mass: $M_{\tilde e}=300,\, 500,\,1000\gev$.
    Here $t_0=1\,$s and the neutralino is again pure bino. Our choice
    of Eq.~(\ref{energy}) is represented by the black vertical line.}}
\label{energy-dep}
\end{figure}

\begin{figure}[ht!]
\vspace{9.26cm}
\includegraphics{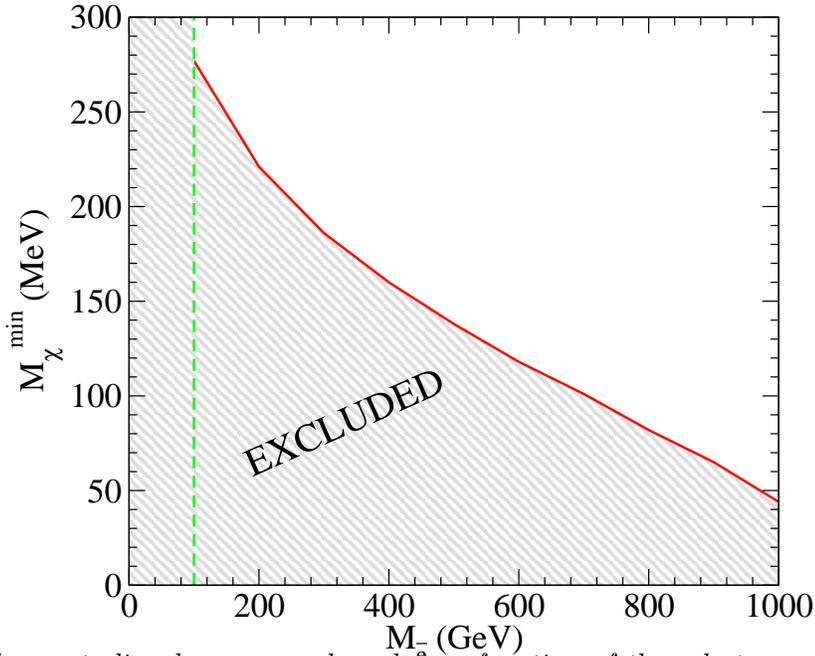}
\caption{\it The neutralino lower mass bound as a function
  of the selectron mass $M_{\tilde e}$, when the Raffelt criterion is
  applied to the emissivity for electron-positron annihilation to
  neutralinos.  The vertical line indicates the lower bound on the
  selectron mass from LEP2~\protect{\cite{adlo}}.}
\label{ee-raffelt}
\end{figure}

Once we have computed $E_\lsp(M_\lsp)$ we can determine a smallest
permitted neutralino mass by requiring the emitted energy to be below
$\emax$ of Eq.~(\ref{energy}).  If we choose $\emax=10^{52}\,$erg,
$t_0=1\,$s and the neutralino to be pure bino then for $M_{\tilde
e}=300\gev$ we have:
\begin{equation}
M_\lsp^{\rm min}=243\mev,
\end{equation}
while for $M_{\tilde e}=1$~TeV we find $M_\lsp^{\rm min}=90\mev$.  For
a massless neutralino to be allowed we require $M_{\tilde e} \geq
1275\gev$.

The value of $M_\lsp^{\rm min}$ is shown for a wide range of selectron
masses in Fig.~\ref{selectron}.  Note that values of $M_\lsp$ below
the solid (red) line are forbidden, as are values of $M_{\tilde e}$ to
the left of the solid line, since in either case the total energy
produced by the process (\ref{Xs-e2}) in the first one second will be
larger than $\emax$.  The lower selectron mass bound from
LEP2~\cite{adlo}:
\begin{equation}
M_{{\tilde e}}\; >\; 99.6\gev\,,
\end{equation}
is also indicated by the vertical dashed (green) line in the figure.

How sensitive are these results to the choice $t_0=1$ second? We made
this choice because we expect most of the neutralino power to be
emitted in a burst during the early, hottest, part of the supernova.
This is indeed the case. If we integrate the power out further,
thereby increasing $t_0$, and again apply the criterion
(\ref{energy}), we obtain the bounds shown in Fig.~\ref{time-dep}.
Once $t_0 > 1$ second the bound is essentially independent of $t_0$.

As discussed above, the exact value chosen for $\emax$---$10^{52}$
ergs---is somewhat arbitrary.  In Fig.~\ref{energy-dep} we show the
dependence of the lowest-allowed $M_\lsp$ on the choice of $\emax$.
This is done for several values of the selectron mass. Note that if
$\emax$ is decreased (increased) by a factor of two then $M_\lsp^{\rm
  min}$ becomes at most 25\% larger (40\% smaller).

Thus, the main dependence of the neutralino mass bound is on the
selectron mass, which sets the effective electron-neutralino coupling
strength.  In addition, it is important to note that most of the
neutralino production from electron-positron annihilation occurs in
the outermost 10\% of the star ($\frac{9}{10}R_c\lsim r\lsim
R_c$). Looking at the profiles used for density and temperature during
the first one second~\cite{Burrows:me} the reason for this becomes
apparent. At early times, this outer region has the highest
temperature, and, more importantly, the lowest electron degeneracy,
$\eta$. Since the rate of the process (\ref{Xs-e2}) is suppressed by
factors of $e^{-\eta}$, the smaller values of $\eta$ near the surface
of the star mean that the majority of neutralinos are produced
there. This will prove critical when we look at the effect of
neutralino trapping on our bounds in Sec.~\ref{trap}.

\subsection{Raffelt Criterion for $e^+e^-\ra\lsp\lsp$}

Although we have used the best supernova input available for radial
temperature and density profiles, and have argued that our results are
not especially sensitive to our choices for $\emax$ and $t_0$, we would
like to check the bounds on $M_\lsp$ as a function of $M_{\tilde{e}}$
that resulted from our modeling of the supernova.
To this end we now turn to the Raffelt criterion (\ref{raffelt}). This is
an estimate of what would happen were the neutralinos implemented in a full
simulation of SN1987A. It is a test that is local in space and time---unlike
the integral measure (\ref{energy}).

The one parameter we are free to choose in applying (\ref{raffelt}) is
the temperature at which $\dot{\cal E}_\lsp$ is to be computed.
Previous work suggests $T_c=30\mev$ is a reasonable choice
\cite{Burrows:1988ah,HPPR,HPRS}. Computing the emissivity due to $e^+
e^- \ra \lsp\lsp$ at this temperature, and at a density of $3 \times
10^{14}~{\rm g}/{\rm cm^3}$, produces the constraint on neutralino and
selectron masses shown in Fig.~\ref{ee-raffelt}. The solid (red) line
indicates the neutralino and selectron masses for which the emissivity
of the process (\ref{Xs-e2}) is exactly $10^{19}$ ergs/g/s.
Neutralino masses below the solid (red) line, and selectron masses to
the left of it, are forbidden. According to the criterion
(\ref{raffelt}), if the supersymmetric particles had such masses the
time-structure of the observed SN1987A neutrino pulse would have been
noticeably different.

Note that the bound obtained with the Raffelt criterion is a little
less stringent than that obtained from (\ref{energy}) and depicted in
Fig.~\ref{selectron}. The two criteria would be in agreement if we
chose $\emax$ to be roughly a factor of two larger. Alternatively, if
in the Raffelt criterion we chose the supernova core temperature to be
$T_c=34\mev$, we would also obtain agreement between the two
approaches.  

Thus the qualitative agreement between these ``global'' and ``local''
criteria gives us confidence in the computationally-simpler Raffelt
criterion---confidence that is supported by detailed simulations of
the impact of axions~\cite{Burrows:1988ah} and
KK-gravitons~\cite{HPPR} on the SN1987A neutrino signal. From now on
we will use the Raffelt criterion to assess the impact of neutralinos
on the cooling of SN1987A.

%
%
\section{Neutralino-Nucleon Interactions}

\label{nn-eff}

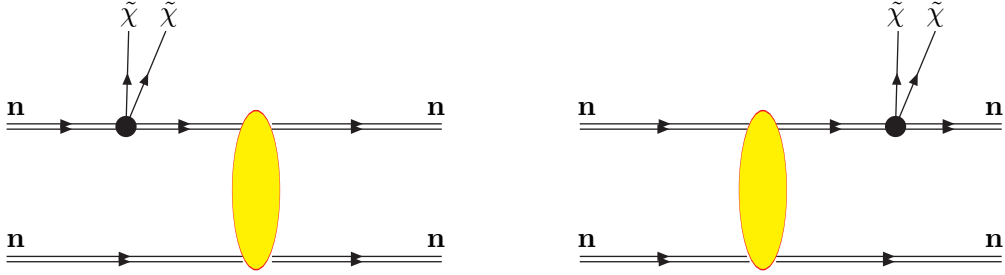
\begin{figure}[t]
\begin{picture}(300,110)(-20,0)
\ArrowLine(1,70)(46,70)
\ArrowLine(46,70)(90,70)
\ArrowLine(1,68)(46,68)
\ArrowLine(46,68)(90,68)
\Vertex(46,69){4}
\ArrowLine(46,69)(47,105)
\ArrowLine(46,69)(61,105)
\ArrowLine(1,20)(90,20)
\ArrowLine(1,18)(90,18)
\ArrowLine(101,20)(165,20)
\ArrowLine(101,18)(165,18)
\ArrowLine(101,70)(165,70)
\ArrowLine(101,68)(165,68)
\COval(95,45)(30,9)(0){Red}{Yellow}
\Text(1,76)[l]{${\rm\bf n}$}
\Text(1,26)[l]{${\rm\bf n}$}
\Text(160,76)[l]{${\rm\bf n}$}
\Text(160,26)[l]{${\rm\bf n}$}
\Text(44,112)[l]{${\lsp}$}
\Text(59,112)[l]{${\lsp}$}
\ArrowLine(217,70)(281,70)
\ArrowLine(217,68)(280,68)
\ArrowLine(217,20)(280,20)
\ArrowLine(217,18)(281,18)
\ArrowLine(291,20)(376,20)
\ArrowLine(291,18)(376,18)
\ArrowLine(291,68)(336,68)
\ArrowLine(291,70)(336,70)
\ArrowLine(336,68)(376,68)
\ArrowLine(336,70)(376,70)
\Vertex(336,69){4}
\ArrowLine(336,69)(337,105)
\ArrowLine(336,69)(351,105)
\COval(286,45)(30,9)(0){Red}{Yellow}
\Text(217,76)[l]{${\rm\bf n}$}
\Text(217,26)[l]{${\rm\bf n}$}
\Text(371,76)[l]{${\rm\bf n}$}
\Text(371,26)[l]{${\rm\bf n}$}
\Text(334,112)[l]{${\lsp}$}
\Text(349,112)[l]{${\lsp}$}
\end{picture}
\caption{{\it Two diagrams contributing to neutralinostrahlung from
the nucleon-nucleon system.  Time flows from left to right. The
nucleons are denoted by double lines. The (yellow) oval indicates the
nucleon-nucleon interaction. The black circle indicates the effective
neutralino-nucleon coupling which we compute in
Sec.~\ref{nn-eff}. Two diagrams which are the same, except
that the neutralinos are radiated from the bottom lines, are not shown.}}
\label{neutralinostrahlung}
\end{figure}

We now turn our attention to the ``binostrahlung'' process
(\ref{Xs-n}). We wish to evaluate the emissivity due to this
reaction. To do this, in Sec.~\ref{nn} we compute the Feynman
graphs shown in Fig.~\ref{neutralinostrahlung}. This gives the
amplitude for the process (\ref{Xs-n}) to leading-order in the
soft-radiation approximation.

These two diagrams, together with their partners under $1\lra 2$
interchange, give the dominant result for binostrahlung in the limit
that the total neutralino energy $\omega\ra 0$, since they are the
only graphs which diverge ($\sim 1/\omega$) in that limit.  Indeed,
were the radiation a single photon, rather than a $\lsp \lsp$ pair,
the graphs in Fig.~\ref{neutralinostrahlung} would yield the
leading-order Low energy theorem for $NN\ra NN\gamma$
\cite{HPR,AD66,Low}.  In order to evaluate them we need to know
two amplitudes: the nucleon-nucleon interaction, denoted by a (yellow)
oval in Fig.~\ref{neutralinostrahlung}, and the effective
neutralino-nucleon interaction, represented by a large black dot in
Fig.~\ref{neutralinostrahlung}.

In the soft-radiation limit the intermediate-state nucleon line in
Fig.~\ref{neutralinostrahlung} is almost on-shell.  Thus, as we
explain in Sec.~\ref{nn} and Appendix A below, to leading order in the
soft-radiation expansion, the nucleon-nucleon interaction occurs
on-mass-shell. In consequence, it can be reconstructed from $NN$
scattering data, giving a model-independent result for this piece of
input to the diagrams in Fig.~\ref{neutralinostrahlung}.

In this section, we focus on the other piece of input to these
diagrams: the neutralino-nucleon coupling. We will show that in the
limit of long-wavelength (i.e. soft) neutralinos this coupling is
expressible in terms of the spin content of the nucleon, $\Delta
q$. We evaluate the $\lsp n$ coupling using two different sets of
results for $\Delta u$, $\Delta d$, and $\Delta s$: the predictions of
the non-relativistic quark model (NRQM), and those of the quark-parton
picture at leading order in $\alpha_s$ (LO-QPM). The
neutralino-neutron coupling turns out to be very different in these
two cases, and this difference has a considerable impact on our final
predictions for the emissivity due to $nn$ binostrahlung. We believe
this gives an estimarte of the range of permitted values.

\subsection{The Bino-Quark Four-Fermion Effective Lagrangian}

\begin{figure}[t!]
\begin{picture}(300,110)(0,0)
\COval(45,45)(36,9)(0){Red}{Green}
\Line(10,70)(38.1,70)
\Line(10,48)(35.3,48)
\Line(10,36)(35.3,36)
\Line(10,24)(37.2,24)
\Line(53,70)(75,70)
\Line(54,48)(172,48)
\Line(54,36)(172,36)
\Line(53,24)(172,24)
\Line(75,70)(110,100)
\DashLine(75,70)(130,70){3}
\Line(130,70)(165,100)
\Line(130,70)(170,70)
\Text(88,107)[l]{$\lsp(k_1)$}
\Text(165,107)[l]{${\lsp}(k_2)$}
\Text(110,78)[l]{$\rm \bf q $}
\Text(113,81)[l]{$\tilde{}$}
\Text(170,78)[l]{$\rm\bf q$}
\Text(61,78)[l]{${\rm\bf q}$}
\Text(10,78)[l]{${\rm \bf q}$}
%
%
\COval(285,45)(36,9)(0){Red}{Green}
\Line(250,70)(278.1,70)
\Line(250,48)(275.3,48)
\Line(250,36)(275.3,36)
\Line(250,24)(278.1,24)
\Line(293,70)(315,70)
\Line(293,48)(410,48)
\Line(293,36)(410,36)
\Line(293,24)(410,24)
\Line(315,70)(405,110)
\DashLine(315,70)(380,70){3}
\Line(380,70)(350,110)
\Line(380,70)(410,70)
\Text(325,112)[l]{$\lsp(k_1)$}
\Text(408,112)[l]{$\lsp(k_2)$}
\Text(352,77)[l]{$\rm \bf q$}
\Text(355,80)[l]{$\tilde{}$}
\Text(410,78)[l]{$\rm \bf q$}
\Text(301,78)[l]{$ \rm \bf q$}
\Text(250,78)[l]{$ \rm \bf q$}
\end{picture}
\caption{{\it The two diagrams contributing to neutralinostrahlung from
a nucleon. The (green) oval denotes the nucleon bound state.  The
horizontal lines are the partons with ${\rm \bf q=u,d,s}$. The virtual
squark ${\rm \bf \tilde{q}}$ can be either left- or right-handed.}}
\label{fig-strahl}
\end{figure}
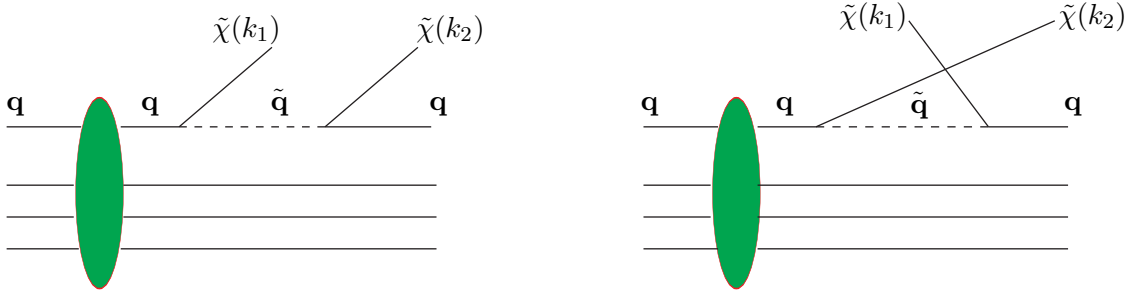

The Lagrangian for a pure bino-neutralino coupling to a quark and a
squark is given by \cite{gunion}:
\begin{eqnarray}
\hspace{-0.5cm}{\cal L}_{q\tilde q \lsp} & = & -\,
\frac{e}{\sqrt{2}\cw} \sum_q
\left[Y_{q_R}\,({\bar q}_R P_L{\lsp})\,{\tilde q}_R + Y_{q_L}
\,{\tilde q}^*_L ({\bar\lsp}\,P_L q)\,
\right],
\end{eqnarray}
where $e$ denotes the electric charge. The sum runs in our case over
the quarks $q=u,d,s$ and ${\tilde q}_{L,R}$ denotes the corresponding
left- or right-handed squark. The hypercharges $Y_{q_R}$ for the
right-handed singlet quarks are given by: $Y_{u_R}=-\frac{4}{3}$, and
$Y_{d_R}= Y_{s_R}= \frac{2}{3}$. For the left-handed quarks
$Y_{q_L}=\frac{1} {3}$.  For each term in the Lagrangian we obtain two
crossed Feynman diagrams contributing to neutralino pair emission from
a nucleon, as shown in Fig.~\ref{fig-strahl}. In the following
we shall assume that the squark masses are degenerate:
\begin{equation}
M_{\tilde q}\equiv M_{{\tilde d}_L} =M_{{\tilde d}_R}
=M_{{\tilde u}_L}=M_{{\tilde u}_R}\,.
\end{equation}
Since the momentum transfer in the neutralinostrahlung in our case
satisfies $Q^2 \ll M_{\tilde q}^2$, we can obtain an effective
four-fermion operator for the $qq\lsp\lsp$ interaction for a pure bino
\cite{EF88}:
\begin{eqnarray}
{\cal L}_{qq\lsp\lsp} & = & 
\frac{e^2}{8 \cos^2\theta_w\, M^2_{\tilde{q}}} ({\bar\lsp}
 \gamma^{\mu}\gamma^5 \lsp) \sum_q \left[Y_{q_L}^2\,
 ({\bar q}\gamma_{\mu}P_L q)\,
-
Y_{q_R}^2 \,({\bar q}\gamma_{\mu}P_R q)\,\right]. \label{eq:nonrel}
\end{eqnarray}
Here we have performed a Fierz transformation in order to separate the
neutralino current from the quark current.  The vector current for
the neutralinos vanishes since they are Majorana fermions.
The quark current can be split into a vector and an axial-vector part:
\begin{equation}
{\cal L}_{qq\lsp\lsp}= -\frac{e^2}{16 M_{\tilde q}^2 \cos^2 \theta_W}\,
(\bar{\lsp} \gamma^\mu \gamma_5 \lsp)
\sum_q [W_q^2\; \bar{q} \gamma_\mu q + X_q^2 \;\bar{q} \gamma_\mu 
\gamma_5 q]\,,\label{eff-lag}
\end{equation}
with the effective charges ($W_q^2\equiv Y_{q_R}^2-Y_{q_L}^2$,
$X_q^2\equiv Y_{q_R}^2+Y_{q_L}^2$):
\begin{eqnarray}
W_u^2=\;\frac{15}{9}\,; &\qquad& W_d^2=\;W_s^2\;=\;\frac{1}{3}
\,,\\ &&\nonumber \\
X_u^2=\;\frac{17}{9}\,; &\qquad& \,X_d^2 =\;X_s^2\;=\;\frac{5}{9}
\,. 
\end{eqnarray}

\subsection{Nucleon Matrix Elements}
\label{sec-nmes}

The neutralino coupling to nucleons can be calculated by taking matrix
elements of the interaction in Eq.~(\ref{eff-lag}) between nucleon
states.  In the limit that $Q^2 \ll \Lambda_{QCD}^2$ this interaction
energy can be rewritten in terms of an axial neutralino current
coupling to quark vector and axial-vector current matrix elements:
\begin{equation}
V_{\lsp\lsp n n}=\frac{e^2}{16 M_{\tilde q}^2 \cos^2 \theta_W}\,
[\,\bar{\lsp} \gamma^\mu \gamma_5 \lsp]\,
\sum_q (W_q^2 V^{(q)}_\mu + X_q^2 A^{(q)}_\mu)\,,
\end{equation}
with:
\begin{eqnarray}
V^{(q)}_\mu&=&\langle n|\int d^3x \, \bar{q}(x) \gamma_\mu q(x)|n \rangle\,,\\
A^{(q)}_\mu&=&\langle n|\int d^3x \, \bar{q}(x) \gamma_\mu \gamma_5 q(x)
|n \rangle\,.
\label{eq:Adef}
\end{eqnarray}
These quark-current matrix elements simplify to:
\begin{eqnarray}
V^{(q)}_\mu=\langle n|\frac{{\rm P}_{\mu}}{M_n}|n\rangle N_q^{(n)}\,,\\
A^{(q)}_\mu=\langle n|\Sigma_\mu|n \rangle\, \Delta q^{(n)},\label{Ares}
\end{eqnarray}
where ${\rm P}_\mu$ and $M_n$ are the momentum and the mass of the
neutron, $N_q^{(n)}$ is the number of valence quarks of type $q$ in
the neutron, $\Sigma_\mu$ is twice the Pauli-Lubanski spin vector, and
$\Delta q^{(n)}$ is the average contribution of quarks of type $q$ to
the total neutron spin.

For a neutron at rest, the only component of $V^{(q)}_\mu$ which is
non-zero is the zeroth one, while the four-vector $\Sigma_\mu$ has
only non-zero spatial components. Thus we arrive at an interaction
energy:
\begin{equation}
V_{\lsp\lsp nn}=\frac{e^2}{16 M_{\tilde q}^2 \cos^2 \theta_W}
\left[\bar{\lsp} \gamma^0 \gamma_5 \lsp \left(\sum_q W_q^2
N_q^{(n)}\right)n^\dagger n - \bar{\lsp} \vec{\gamma} \gamma_5 \lsp \cdot
(n^\dagger \vec{\sigma}_n n)\left(\sum_q X_q^2\, \Delta
q^{(n)}\right)\right]\,,
\label{V-eff}
\end{equation}
with $\vec{\sigma}_n$ the three-vector of the neutron Pauli spin
matrices.  Corrections to Eq.~(\ref{V-eff}) are suppressed by powers
of ${\bf P}_n/M_n$ and/or powers of $Q R_n$, with $R_n$ the
neutron's size.  This allows us to write an effective four-fermion
Lagrangian for the interaction of neutralinos with non-relativistic
neutrons:
\begin{equation}
{\cal L}_{\lsp\lsp nn}=-\frac{G_{SUSY}}{2 \sqrt{2}} a^\mu 
\left[n^\dagger \left(c_{\rm v}^\chi
\delta_{\mu,0} - c_{\rm a}^\chi \delta_{\mu,i} \sigma_i\right)n\right]\,,
\label{eff-lag-nuc}
\end{equation}
with the axial-vector neutralino current:
\begin{equation}
a^\mu \equiv \bar{\lsp} \gamma^\mu \gamma_5 \lsp\,,
\end{equation}
and the effective couplings:
\begin{eqnarray}
&&\qquad \qquad G_{SUSY} \equiv \frac{e^2}{4 \sqrt{2} \cos^2 \theta_W M_{\tilde q}^2}\,,\\
&& c_{\rm v}^\chi = \frac{19}{9}\,; \qquad
c_{\rm a}^\chi = \frac{17}{9}\, \Delta d + \frac{5}{9} (\,\Delta u + \,\Delta s).
\label{eq:cachi}
\end{eqnarray}
(In (\ref{eq:cachi}) we have used the conventional definitions of the $\Delta
q$'s and isospin.)

The two diagrams in Fig.~\ref{neutralinostrahlung} enter with opposite
signs in the computation of the amplitude for the radiative
process. Thus, as we discuss further in Sec.~\ref{nn}, to leading
order in $1/M_n$ the vector-current interaction does not contribute to
the neutralinostrahlung emissivity.  Hereafter we disregard its
contribution in order to get the lowest-order result for the
binostrahlung emissivity in the soft-radiation approximation.  It is
worth bearing in mind though that $1/M_n$ corrections, as well as
effects due to non-zero proton fraction, could render vector-current
radiation quite significant in binostrahlung.

Note that, in general, the vector
current dominates neutralino-neutron {\it scattering}, which we will
examine in Sec.~\ref{trap}. 

This leaves us with the task of estimating $\Delta u$, $\Delta d$, and
$\Delta s$. We consider only two out of a plethora of approaches:
\begin{enumerate}
\item Employ the non-relativistic quark model (NRQM). In this case:
\begin{equation}
\Delta u=\frac{4}{3}; \qquad
\Delta d=-\frac{1}{3}; \qquad
\Delta s=0,
\end{equation}
giving:
\begin{equation}
c_{\rm a}^\chi=\frac{1}{9}\,.
\end{equation}

\item Employ the leading order quark parton model
  (LO-QPM)~\cite{Filippone:2001ux}. This obeys the constraint $\Delta
  u - \Delta d=g_A$, and does not fare too badly with respect to the
  constraint from the $F/D$ ratio extracted from hyperon
  decays~\cite{Ke97}:
\begin{equation}
\Delta u + \Delta d - 2 \Delta s \approx 0.682\,.
\end{equation}
The values from Ref.~\cite{Filippone:2001ux}  are:
\begin{equation}
\Delta u=0.78 \pm 0.03\,; \qquad \Delta d=-0.48 \pm 0.03\,; 
\qquad \Delta s=-0.14 \pm 0.03\,. \label{best-qpm}
\end{equation}
These values rely on the extraction of the flavour-singlet matrix
element from the zeroth moment of the protons $g_1$ structure
function, and its subsequent evolution to low-$Q^2$.  This evolution
is perhaps questionable, since below $Q^2 \sim 1\gev^2$ it becomes
non-perturbative. Nevertheless, the values (\ref{best-qpm}) give:
\begin{equation}
c_{\rm a}^\chi=-0.55 \pm 0.06\,,
\end{equation}
a result five times as large as the NRQM one, and of opposite
sign.
\end{enumerate}

\subsection{Ratio of Neutralino and Neutrino Processes}

\label{sec-neutrino}

We now follow Ref.~\cite{Ellis:1988aa}, and take the ratio of the
coupling $c_a^\chi$ to the corresponding neutrino coupling, in an
attempt to eliminate some of the uncertainties in the computation.

The effective Lagrangian for the neutrino case is~\cite{FM79,HPR,RS95}:
\begin{equation}
{\cal L}_{\nu\nu nn}=-\frac{G_F}{2 \sqrt{2}} \;l^\mu \;
[n^\dagger (c_{\rm v}^{\nu} 
\delta_{\mu,0} - c_{\rm a}^{\nu} \delta_{\mu,i} \sigma_i) n]\,,
\label{eq:neuteffL}
\end{equation}
with the leptonic current:
\begin{equation}
l^\mu \equiv \bar{\nu} \gamma^\mu (1 - \gamma_5) \nu\,.
\end{equation}
Here we will drop the vector-current piece, since it does not
contribute to $nn \rightarrow nn \nu \bar{\nu}$. Meanwhile,
arguments analogous to those of the previous subsection yield:
\begin{equation}
c_{\rm a}^\nu=-[\Delta u - \Delta d + \Delta s].
\end{equation}

In the non-relativistic quark model we have:
\begin{equation}
c_{\rm a}^\nu=-\frac{5}{3}\,,
\label{eq:QMca}
\end{equation}
while using the values for the $\Delta q\,$'s from the quark-parton
model in Eq.~(\ref{best-qpm}) we find:
\begin{equation}
c_{\rm a}^\nu=-1.12 \pm 0.05\,.
\end{equation}
The latter value is about 33\% lower than in the NRQM: a significantly
smaller discrepancy than in the neutralino case.

Therefore for our chosen evaluation procedures the ratio of neutralino
to neutrino axial coupling constants is:
\begin{equation}
\frac{c_{\rm a}^\chi}{c_{\rm a}^\nu}=\left\{\begin{array}{ll}
                                -\frac{1}{15}  & \quad \mbox{NRQM}\,,\\ &\\
                                0.49 \pm 0.08 & \quad \mbox{LO-QPM}\,.
                                \end{array} \right.
\end{equation}

In order to use the the effective Lagrangian (\ref{eq:neuteffL}) to
compute the neutrino emissivity due to the reaction $nn \rightarrow nn
\nu\bar {\nu}$, we require the spin-summed matrix element for
$nn \rightarrow nn \nu \bar{\nu}$~\cite{FM79,HPR,RS95}:
\begin{equation}
\sum_{\nu~{\rm spin}} |{\cal M}|^2=\frac{G_F^2 (c_{\rm a}^\nu)^2}{8}
{\rm Tr}(l_i l_j)\left(\frac{16}{\omega^2}\right) {\cal H}_{ij}\,.
\label{forneutrinos}
\end{equation}
Here ${\cal H}_{ij}$ is the hadronic tensor, which corresponds to two
insertions of the neutron spin in the hadronic (in this case $nn$)
state.  The evaluation of this quantity will be discussed in
Sec.~\ref{nn}, but note that we have already used the fact that only
its spatial components are non-zero in the non-relativistic limit.
(Here and below we denote spatial indices by Roman letters.)  Also,
for later convenience, we have removed a factor of $1/\omega^2$, to
take care of the leading behavior of $|{\cal M}|^2$ in the
soft-radiation limit, as well as a factor of $16$, to account for
identical particles and the fact that the neutrinos couple to both
nucleons. ($\omega$ is the total emitted neutralino energy.) The sum
over hadronic states in ${\cal H}_{ij}$ will then only include states
permitted by the Pauli principle.

Meanwhile, ${\rm Tr}(l_i l_j)$ are the space-space
components of the neutrino tensor:
\begin{eqnarray}
\ell^{\mu \nu} &\equiv& {\rm Tr}[\not \! k_1
\gamma_\mu (1 - \gamma_5) \not \! k_2 \gamma_\nu (1 - \gamma_5)]\,,\\
&=&8\left[k_1^\mu k_2^\nu + k_2^\mu k_1^\nu 
- g^{\mu \nu} k_1 \cdot k_2 + i \epsilon_{\alpha \mu \beta \nu} k_1^\alpha
k_2^\beta \right],
\end{eqnarray}
in agreement with Friman and Maxwell~\cite{FM79} ($k_{1,2}$ are the
outgoing neutrino momenta).

Now, in order to compute the spin-squared matrix element for $nn
\rightarrow nn \lsp \lsp$ we proceed by analogy with this neutrino
calculation. The spin-summed matrix element for neutralinostrahlung
which is integrated to yield the neutralino emissivity due to the
process (\ref{Xs-n}) can be written as:
\begin{equation}
\sum_{\lsp~{\rm spin}} |{\cal M}|^2=\frac{G_{SUSY}^2 {c_{\rm a}^\chi}^2}{8}
{\rm Tr}(a_i a_j)\left(\frac{16}{\omega^2}\right) {\cal H}_{ij}\,.
\label{hijdef}
\end{equation}
Crucially, the hadronic tensor here is the same as the one in
Eq.~(\ref{forneutrinos}), since neutralino-pair production is also
induced by two insertions of the neutron-spin operator in the hadronic
system.

We now compute the neutralino tensor ${\rm Tr}(a_i a_j)$.  It is to be
understood as the space-space piece of the four-tensor:
\begin{eqnarray}
{\cal N}^{\mu \nu} &\equiv& Tr[(\not \! k_1 + M_\lsp)\gamma^\mu \gamma_5 
(\not \! k_2 - M_\lsp) \gamma^\nu \gamma_5]\,, \\  &=&4 \left[k_1^\mu 
k_2^\nu + k_1^\nu k_2^\mu - (k_1 \cdot k_2 - M_\lsp^2) g^{\mu\nu}\right],
\end{eqnarray}
where $k_1$ and $k_2$ are now the outgoing {\it neutralino} four-momenta.

In order to compare the neutralino and neutrino computations it is
convenient to define $\alpha_\lsp$ to be the ratio of neutralino to
neutrino spin-summed matrix elements, with the ratio taken in the
limit $M_\lsp=0$. Then:
\begin{equation}
\alpha_\lsp\equiv\left(\frac{G_{SUSY} 
\;c_{\rm a}^\chi}{G_F\; c_{\rm a}^\nu}\right)^2
\frac{{\cal N}_{ij} {\cal H}_{ij}}{\ell_{ij} {\cal H}_{ij}},
\label{eq:alphachi}
\end{equation}
where ${\cal H}_{ij}$ is the hadronic spin response in both cases.  

In the computation that follows in Sec.~\ref{nn}, we assume that the
nucleons are in thermal equilibrium, with a common temperature $T_c$.
Thus the nucleon kinetic energies are of order $T_c$, and their
momenta are of order $\sqrt{M_n T_c}$. The centre-of-mass energy of
the reaction (\ref{Xs-n}) is therefore also thermally distributed
around $2T_c$.  This means that neutralinos are produced with a common
``temperature'' $ \sim T_c$.  For $M_\lsp\lsim 50\mev$, the
neutralinos are then relativistic and have momenta of order $T_c\ll
\sqrt{M_n T_c}$.  Consequently, to leading order in $\sqrt{T_c/ M_n}$
we can ignore the recoil of the hadronic system when the neutralinos
are emitted, and ${\cal H}_{\mu \nu}$ will then be independent of
${\bf k}_1$ and ${\bf k}_2$.  This allows us to perform the integral
over these variables and obtain an ``angle-averaged'' neutralino
tensor:
\begin{equation}
{\overline{\cal N}}_{\mu \nu} \equiv \int \frac{d \Omega_1}{4 \pi}
\int \frac{d \Omega_2}{4 \pi} {\cal N}_{\mu \nu}\,,
\label{barndef}
\end{equation}
and analogously for the neutrino case. For the space-space
components---which are all that are relevant here---this gives:
\begin{eqnarray}
\overline{\cal N}_{ij}&=&\delta_{ij} 4 (\omega_1 \omega_2 - M_\lsp^2)\,;
\nonumber\\ \overline\ell_{ij}&=&\delta_{ij} 8 \omega_1 \omega_2\,,
\label{eq:angleavgtensors}
\end{eqnarray}
where $\omega_{1,2}$ are the energies of the two emitted
neutrinos/neutralinos.

In any observable---including emissivities---in which the outgoing
neutr(al)inos are not detected we can use these ``angle-averaged''
leptonic tensors in the evaluation of the spin-summed matrix
element---provided, of course, that we truly are in the kinematic
regime where recoil of the hadronic system can be neglected.
Calculating $\alpha_\lsp$ using the tensors (\ref{eq:angleavgtensors})
we get:
\begin{eqnarray}
\alpha_\lsp&=&\frac{1}{2} 
\left(\frac{G_{SUSY}\; c_{\rm a}^\chi}{G_F\; c_{\rm a}^\nu}\right)^2\\
&=&\frac{\sin^4 \theta_W}{2} \left(\frac{M_Z}{M_{\tilde q}}\right)^4 
\left(\frac{c_{\rm a}^\chi}{c_{\rm a}^\nu}\right)^2.
\label{eq:caratios}
\end{eqnarray}
Note that the hadronic tensor has canceled, so this ratio is
independent of the model of the nuclear medium that is adopted. Note
also that all of the {\it nucleon}-model dependence is now isolated in
the last factor.  Finally, Eq.~(\ref{eq:caratios}) gives:
\begin{equation}
\alpha_\lsp \approx \sin^4 \theta_W \left(\frac{M_Z}{M_{\tilde q}}\right)^4 
\times  \left\{ \begin{array}{ll}
                \frac{1}{450}  & \quad \mbox{NRQM}\,,\\ & \\
                \frac{1}{8}      & \quad  \mbox{LO-QPM}\,,
        \end{array} \right.
\label{res-alpha}
\end{equation}
where we have written $0.49 \approx \frac{1}{2}$. $\alpha_\lsp$ is
more than fifty times larger in the LO-QPM than in the NRQM.

\section{Emission from a Neutron Gas}
\label{nn}

We are now ready to compute the emissivity due to neutron-neutron
``neutralinostrahlung'', the process (\ref{Xs-n}).  (For simplicity we
will restrict ourselves to pure neutron matter. The generalization to
arbitrary proton fraction is straightforward, if somewhat cumbersome.)
We shall derive a general emissivity, which is valid for any
particle-pair which couples axially to the nucleons.  We then demand
that this emissivity is not larger than the Raffelt bound
(\ref{raffelt}), and so set bounds on $\alpha_\lsp$. At the end of
this section we use the results of Sec.~\ref{nn-eff} to translate our
bound on $\alpha_\lsp$ into a bound on the squark mass.

It was recently shown that, for soft radiation, the neutrino
emissivity due to $nn \rightarrow nn \nu \bar{\nu}$ can be directly
related to the on-shell nucleon-nucleon data~\cite{HPR}. For
$M_\lsp\lsim50\mev$, this ``soft-radiation approximation'' also
provides a model-independent way to compute the emissivity due to
neutralinostrahlung. On the other hand, for neutralino masses of order
100 MeV or more the radiation cannot be regarded as soft, since
$M_\lsp \sim m_\pi$, and the pion plays an important role in $NN$
scattering and in bremsstrahlung dynamics.  However, even for large
(from a nuclear physics point of view) neutralino masses we still
regard the method presented here as a useful way to get an order of
magnitude estimate of the emissivity due to $nn \rightarrow nn
\lsp\lsp$. As will be demonstrated below, this emissivity depends
sensitively on $M_\lsp$ through the Boltzmann factors and so, even if
our estimate of the matrix element is off by an order of magnitude,
the neutralino-mass bound changes by only about 30\%.

The emissivity for the radiation of neutralinos from an interacting neutron
gas is given by
\begin{eqnarray}
\dot {\cal E}
&=&\int d\omega \int \left[ \prod_{i=1,2}
\frac{d^3 p_i\,d^3p_i'}{(2\pi)^6}\right](2\pi)^4\delta^{(4)}(p_1+p_2-p_1'-p_2'-k)
\nonumber \\
& & \ \qquad \qquad \qquad \qquad \qquad \times \
S \cdot f(E_1)f(E_2)(1-f(E_1'))
(1-f(E_2'))\frac{dE}{d\omega} \ ,
\label{emissform}
\end{eqnarray}
where $S=1/8$ is the symmetry factor taking into account that the
initial and final nucleon pairs, as well as the neutralinos are
identical. $p_1,\,p_2$ are the incoming and $p_1',\,p_2'$ are the
outgoing neutron four-momenta.  $f(E)$ is the Fermi-Dirac distribution
function (\ref{eq:FDdist}).  The energy per unit frequency interval
for a particular two-neutron phase-space element is given by:
\begin{equation}
\frac{dE}{d\omega} = \int \frac{d^3 k_1}{(2 \pi)^3 2 \omega_1}
\frac{d^3 k_2}{(2 \pi)^3 2 \omega_2} \, \omega \, \delta(\omega -
\omega_1 - \omega_2) \delta^{(3)}(k - k_1 - k_2) |{\cal M}|^2.
\label{eq:enloss}
\end{equation}
$\omega$ and ${\bf k}$ are the total energy and three-momentum of the
neutralino pair, and $k_{1,2}$ are the three-momenta of the
neutralinos.  ${\cal M}$ is the matrix element for the process $nn
\rightarrow nn \lsp \lsp$.

To proceed, we now use the decomposition of $|{\cal M}|^2$ into a
hadronic tensor and a neutralino tensor as given in
Eq. (\ref{hijdef}). Introducing this into Eq. (\ref{eq:enloss}) and
using the angle-averaged expression of Eq. (\ref{barndef}) for the
neutralino tensor, we derive
\begin{equation}
  \frac{dE}{d\omega} = \alpha_\lsp G_F^2 {c_{\rm a}^\nu}^2 \frac{1}{
    \pi^4} \int \, d\omega_1 d\omega_2 \,\delta(\omega - \omega_1
  -\omega_2) \, \frac{k_1 k_2}{\omega} \, (\omega_1 \omega_2 - M_\lsp^2)
 {\cal H}_{ii} \, ,
\label{eq:desire}
\end{equation}
Here $\alpha_\lsp,\,G_F,\,c_{\rm a}^\nu$ are as defined in
Sec.~\ref{nn-eff}.  We will now set bounds on $\alpha_\lsp$ which
apply to any axially-coupled light-particle-pairs emitted from
the neutron gas.  We then translate to bounds on the squark mass $M_{\tilde
q}$ using Eq.~(\ref{res-alpha}).

\subsection{Evaluating the Hadronic Tensor}

In Section~\ref{nn-eff}, we saw that when calculating the neutr(al)ino
emissivity only the space-space diagonal elements of the hadronic
tensor are required. To leading order in the soft-radiation
approximation, and provided many-body effects are not large, only the
diagrams shown in Fig.~\ref{neutralinostrahlung} contribute to ${\cal
H}_{ii}$. The tensor can then be expressed in terms of the commutator
of the production operator with the $NN$ scattering matrix
$T_{NN}$~\cite{HPR}.  According to ${\cal L}_{\lsp\lsp nn}$ of
Eq.~(\ref{eff-lag-nuc}), in the non-relativistic limit only two
operator structures govern the $\lsp \lsp$ pair's coupling: the unit
operator and the spin operator. The commutator of the unit operator
with $T_{NN}$ vanishes, and therefore the vector current in
Eq.~(\ref{eff-lag-nuc}) does not contribute to $nn \rightarrow nn \lsp
\lsp$. Thus, at leading order the only structure which contributes to
${\cal H}_{ij}$ is the two-nucleon spin operator, $S_i$.  For a given
two-nucleon state of total energy $E$, we find that the relevant,
diagonal, components of ${\cal H}_{ij}$ are:
\begin{equation}
{\cal H}_{ii}(E)=\sum_{m_s,m_{s}'}|\langle 1\, m_{s}', \, {\bf p}'|\,
[S_i,T_{NN}]\,|{\bf p}, \, 1 \,m_{s} \rangle |^2\,,
\label{eq:hiidef}
\end{equation}
where $|{\bf p}, \, 1\, m_s \rangle$ is a two-nucleon state of total
spin 1, total spin projection $m_s$, and relative momentum ${\bf p}$.
Here, because the binos radiated in Fig.~\ref{neutralinostrahlung}
are soft, the initial and final $nn$ relative momenta obey:
\begin{equation}
{\bf p}^2={\bf p}'^2=ME\,.
\end{equation}
The key point here is that the $NN$ matrix elements in
Eq.~(\ref{eq:hiidef}) are on-shell, and so can be obtained from
nucleon-nucleon scattering data. So, to leading order in the
soft-radiation approximation, there is a direct connection between
this data and the neutralino emissivity $d {\cal E}/dt$: a connection
given by Eqs.~(\ref{eq:hiidef}), (\ref{eq:desire}) and
(\ref{emissform}).

To further simplify Eq.~(\ref{emissform}), spherical symmetry and
energy-momentum conservation can be exploited, thereby eliminating
nine of the eighteen integrals.  We define total, initial relative,
and final relative momenta ${\bf P}$, ${\bf p}$, and ${\bf p}'$,
respectively, and also move to dimensionless
variables~\cite{Brinkmann:1988vi}:
\begin{equation}
\tau ^2=\frac{{\bf P}^2}{8 M T_c}\,, \qquad \delta=\frac{{\bf p}^2-{\bf
p}'\, ^2}{2 M T_c}\,, \qquad \sigma =\frac{{\bf p}^2+{\bf p}'^2}{2MT_c}\,,
\end{equation}
and
\begin{equation}
\eta_n = \frac{\mu}{T_c}\,, \qquad \hat M_\lsp = \frac{M_\lsp}{T_c}\,,
\label{ymhat}
\end{equation}
where $T_c$ is the temperature of the neutron gas in the supernova.
At large temperatures, the angular dependence of the Fermi functions
becomes weak---in the non-degenerate limit the angular dependence
disappears completely. We have checked that at the temperatures we are
studying replacing the product of Fermi functions by their angular
average is justified.  These angular integrals can be performed
analytically \cite{Brinkmann:1988vi}:
\begin{eqnarray}
\int_{-1}^1 \, d\gamma \, f(E_1)f(E_2) =
\frac{e^{-\xi^2}}{\tau (\delta+\sigma) \sinh (\xi^2)}
\ln \left( \frac{\cosh \left[\frac{1}{2}\left\{\xi^2+\tau (\delta+\sigma)
\right\}\right]} {\cosh \left[\frac{1}{2}\left\{\xi^2-\tau 
(\delta+\sigma)\right\}\right]} 
\right) \ ,
\end{eqnarray}
where $\xi^2 = \frac{1}{4}(\delta+\sigma)^2+\tau^2-\eta_n$.  
Doing the angular integrals this way necessitates the introduction of
the angle-averaged hadronic tensor:
\begin{equation}
\overline{\cal H}_{ii} = \int \, \frac{d\hat{p} \, d\hat{p}'}{16
\pi^2} \, {\cal H}_{ii} \ .
\label{hbardef}
\end{equation}
The replacement ${\cal H}_{ii} \longrightarrow \overline{\cal H}_{ii}$
is not only a good approximation, it is a useful one, since the
integrals in Eq.~(\ref{hbardef}) be carried out analytically, when the
$NN$ $T$--matrix is given in partial wave form.  The procedure used to
construct $\overline{\cal H}_{ii}$ from a given set of $NN$ phase
shifts is sketched in Appendix A. The results that follow were
generated using the VPI phase-shift analysis in that
construction~\cite{VPI}.

\subsection{Emissivity}

We now return to the computation of the emissivity
Eq.~(\ref{emissform}). Doing the angular integrals as described leaves us
with a four-dimensional integral which we decompose as
\begin{eqnarray}
\dot{\cal E}=
 \alpha_\lsp\,\frac{G_F^2c_A^2}{8(2\pi)^{11}}
 \,(2MT_c)^{9/2}M_\lsp^4 \int_{\hat M_\lsp}^\infty d\delta 
\, e^{-\delta}h(\delta/ \hat M_\lsp)
\int_\delta^\infty d\sigma \, {\cal F}(\delta,\sigma,\eta_n)\,
 \bar {\cal H}_{ii}(T_c\,\sigma)\,,
\label{eq:final}
\end{eqnarray}
where
\begin{equation}
h(y) =  \frac{1}{y}
 (y - 1)^3
\left[(y^2+1)I_0(y)
-(y-1)^2I_2(y) \right],
\end{equation}
with
\begin{equation}
I_k(y)=\int_{-1}^1 \, dx \, x^k\sqrt{(1-x^2) (Y^2(y)-x^2)}\,,
\end{equation}
and $Y(y)\equiv (y+1)/(y-1)$.  The result of the angular integration of
the Fermi functions is
\begin{equation}
{\cal F}(\delta ,\sigma,\eta_n) = \int_0^\infty d \tau \,
{\cal R}\left(\sqrt{\frac{1}{2}(\sigma + \delta)},\tau,\eta_n \right)
{\cal R}\left(\sqrt{\frac{1}{2}(\sigma - \delta)},\tau,\eta_n \right),
\end{equation}
where
\begin{equation}
{\cal R}(a,\tau,\eta_n)=\frac{1}{\sinh ( a^2+\tau ^2-\eta_n)}
\ln \left(\frac{\cosh \left(\frac{1}{2}((\tau + a)^2-\eta_n)\right)}
{\cosh \left(\frac{1}{2}((\tau - a)^2-\eta_n)\right)}\right).
\end{equation}

\begin{figure}[t]
\vspace{8cm}
\includegraphics{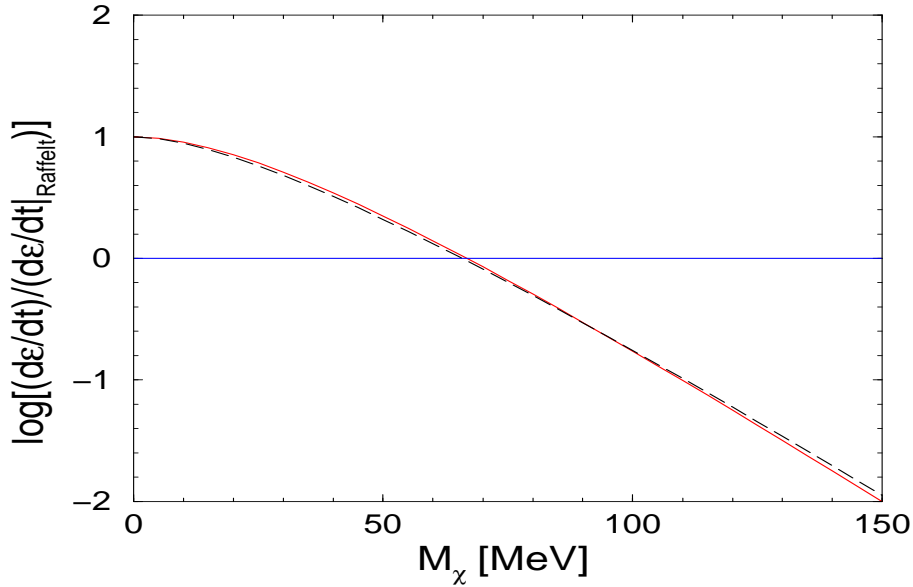}
\vspace*{-1cm}
\caption{\it{Neutralino emissivity in units of $10^{19}$ ergs/g/s as a
    function of the neutralino mass $M_\lsp$. The red solid line shows
    the full result for $\alpha_\lsp=10\, \alpha_\lsp^{\rm crit}$,
    whereas the black dashed line shows the fit according to Eq.
    (\protect{\ref{fit}}). The horizontal blue solid line shows the
    emissivity $10^{19}$ ergs/g/s that is used when the Raffelt
    criterion is applied.}}
\label{mdep}
\vspace{-0.3cm}
\end{figure}

The functions $I_k(y)$ can be evaluated to very high accuracy if we
replace the second factor in the square root by its Taylor expansion
up to third order in $(x/Y)^2$.  Formally one would expect this
approximation to work for $Y \gg 1$ only.  However, even for $Y=1$
both $I_2$ as well as $I_0$ turn out to be within 5\% of the exact
result.  The remaining 3 integrals are evaluated numerically using the
Gauss-Legendre method. The computation was performed for nuclear
matter density and $T_c=30\mev$. The result of the numerical
integration is shown as the solid line in Fig.~\ref{mdep} for
$\alpha_\lsp=2.5 \times 10^{-4}$.
The numerical result is well fit by:
\begin{equation}
\dot{\cal E}
 = 4 \alpha_\lsp \ (1+2\hat M_\lsp+ 0.55  \hat
M_\lsp^{2}) \ e^{-2 \hat M_\lsp} \
 \times \ 10^{23} \
\mbox{[ergs/g/s]} ,
\label{fit}
\end{equation}
where $\hat M_\lsp \equiv M_\lsp/T_c$ as defined in Eq. (\ref{ymhat}).
The fit is good for $M_\lsp \lsim 200\mev$.  Our result represented in
the form (\ref{fit}) allows us to directly derive a neutralino
mass bound for any given coupling strength $\alpha_\lsp$ using the
Raffelt criterion (\ref{raffelt}). Alternatively, we can think of
defining a function $\alpha_\lsp^{\rm crit}(M_\lsp)$ from
Eq.~(\ref{fit}): the coupling at which the Raffelt criterion is
exactly met for a given neutralino mass $M_\lsp$.

\begin{figure}[t]
\vspace{8cm}
\includegraphics{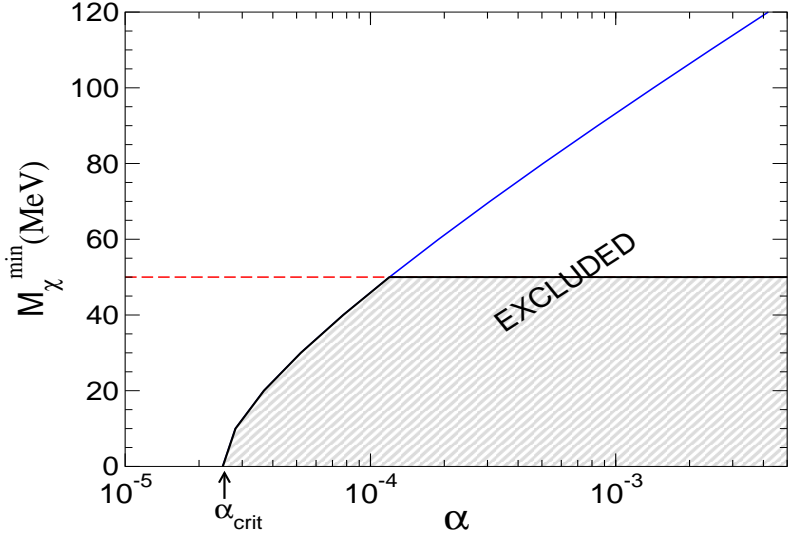}
\caption{\it{The region of the $\alpha_\lsp$-$M_\lsp$(MeV) plane that is
    excluded by our soft-radiation calculation, when applying the
    Raffelt criterion (at $T_c=30\mev$) is indicated by the grey
    shading.  The blue solid line is the curve that defines the values
    of $\alpha_\lsp$ and $M_\lsp$ which produce a neutralino
    emissivity of exactly $10^{19}$ ergs/g/s.  The horizontal (red)
    dashed line indicates the uppermost value of $M_\lsp$ at which the
    soft-radiation approximation should be trusted. }}
\label{fig-alpham}
\end{figure}

By setting $\hat M_\lsp=0$ in Eq.~(\ref{fit}) above, \ie\ a
massless neutralino, we obtain the minimal coupling $\alpha_\lsp^{\rm
  crit}(0)$, for which we can set a bound.  Using Eq.~(\ref{raffelt}) we
obtain
\begin{equation}
\alpha_\lsp^{\rm crit}(0)=2.5 \times 10^{-5}.
\label{ac}
\end{equation}
For $M_\lsp=T_c=30\mev$, $\alpha_\lsp^{\rm crit}(M_\lsp)$ increases to $5.2
\times10^{-5}$. For $M_\lsp>T_c$, the minimum $\alpha_\lsp$ we are
sensitive to changes rapidly with the neutralino mass due to the
exponential dependence seen in Eq.~(\ref{fit}). Using the Raffelt
criterion we can exclude a region in the $\alpha_ \lsp$--$M_\lsp$
plane which is to the right of the curve defined by the function
$\alpha_\lsp^{\rm crit}(M_\lsp)$; this region is shown in
Fig.~\ref{fig-alpham}.

As stated in the beginning of this section, we believe the
soft-radiation approximation we make in our computation is under
control for $M_\lsp\lsim50\mev$. This limiting mass value is indicated
by the horizontal (red) dashed line in Fig.~\ref{fig-alpham}. 
At this
mass value, the smallest $\alpha_\lsp$ value we are sensitive to is
\begin{equation}
\alpha_\lsp^{\rm crit}(50\mev)= 1.2 \times 10^{-4}.
\end{equation}
For $\alpha_\lsp \geq \alpha_\lsp^{\rm crit}(50\mev)$, we can still use the
soft-radiation calculation to obtain a reliable bound, as long as
$M_\lsp \lsim 50$ MeV. This excludes the shaded area in
Fig.~\ref{fig-alpham}.

For larger neutralino masses, the soft-radiation approximation breaks
down, since the emitted energy is of order 100 MeV.  
An approximate bound for $M_\lsp\gsim50\mev$
can still be derived, because most of the dependence of the emissivity
comes from the phase-space integrals in Eq.~(\ref{emissform}), and so
using the form (\ref{hijdef}) to fix the $nn \rightarrow nn \lsp \lsp$
matrix element should give results accurate to an order of
magnitude, even if $M_\lsp$ is as large as 200 MeV.  Thus we argue
that the entire region of $\alpha_\lsp-M_\lsp$ parameter space to the
right of the solid (blue) curve in Fig.~\ref{fig-alpham} is also
excluded by the arguments of this section, although less rigorously so
than the grey shaded area.

\subsection{Bound on Squark Masses}

Using Eq.~(\ref{res-alpha}) to relate $\alpha_\lsp$ to $M_{\tilde{q}}$
we can turn the bounds on $\alpha_\lsp$ we have obtained
into bounds on the squark mass. 
\begin{equation}
M_{\tilde q} \;>\; \left(\frac{2.5 \times 10^{-5}}{\alpha_\lsp^{\rm
                      crit}(M_\lsp)}\right)^{1/4} \;\times
                      \left\{\begin{array}{ll} 132\gev\,, & \quad
                      \mbox{NRQM}\,,\\ &\\ 360\gev\,,& \quad
                      \mbox{LO-QPM}\,,  \end{array} \right.
\label{eq:sqbd}
\end{equation}
where all the dependence on $M_\lsp$ is now in the value of
$\alpha_\lsp^{\rm crit}$, which can be easily obtained from
Eq.~(\ref{fit}) and the Raffelt criterion. 

Eq.~(\ref{eq:sqbd}) is significantly weaker than the bounds found in
Ref.~\cite{Ellis:1988aa}.  There are several reasons for this:
\begin{enumerate}
\item As noted in the Introduction, Ellis {\it et al.} use a
supernova core temperature of $T_c=70\mev$---more than a factor of
two larger than the one employed here. Since the emissivity
(\ref{fit}) varies as a high power of the temperature this is
a crucial difference.

\item The axial hadronic response to which the neutralinos couple is
about four times larger in the Ellis {\it et al.} result than in our
improved computation, because Ref.~\cite{Ellis:1988aa} (effectively)
employed the single-pion-exchange approximation to compute this
response.

\item Ellis {\it et al.} assumed a neutralino which was pure
photino. The effective coupling for a pure-photino neutralino compared
to a pure-bino neutralino is a factor
\begin{equation}
\frac{\alpha_{\tilde\gamma}}{\alpha_\lsp}=256 \, \cos^4\theta_w \,
\left(
\frac{\frac{4}{9}\Delta d + \frac{1}{9}\Delta u + \frac{1}{9}\Delta s}
{\frac{17}{9}\Delta d + \frac{5}{9}\Delta u + \frac{5}{9}\Delta s}
\right)^2\approx 10.28\,
\label{rat-alph}
\end{equation}
larger. (Here we computed $\alpha_{\tilde\gamma}$ analogously to the
manner used to find $\alpha_\lsp$ in Eq.~(\ref{res-alpha}) and used
the LO-QPM values for the $\Delta q$'s.)
\end{enumerate}

Our bounds on the squark mass are thus significantly less restrictive
than those we found for the selectron mass in Sec.~\ref{ep}.

\section{Neutralino trapping}

\label{trap}
\subsection{Diffusive Trapping}

The arguments of this paper apply only if neutralinos free-stream out
of the supernova. In order to determine whether this occurs we must
estimate their mean-free path
\begin{equation}
\lam_\lsp=\frac{1}{n\,\sigma_\lsp}\;,
\end{equation}
via the scattering processes given in
Eqs.~(\ref{scatter1},\ref{scatter2}). Here $n$ is the target particle
number density in the supernova. For the electrons, we take $n_e=8.7
\times 10^{43}~{\rm m}^{-3}$ and for the neutrons $n_n=5.7\times
10^{43}~{\rm m}^{-3}$ \cite{Burrows:me}.  In both cases we approximate
the density to be independent of radius.

We shall first discuss the scattering on electrons and then estimate
the case for scattering on nucleons. In both cases we use the optical
depth criterion~\cite{Burrows:me}:
\begin{equation}
\int_{r_0}^{R_c} \frac{dr}{\lam_\lsp(r)} \leq \frac{2}{3}\,,
\label{eq:optdepth}
\end{equation}
to determine whether neutralinos produced at a depth $r_0$,
free-stream out of the supernova or not. 

Any neutralino propagating out of the proto-neutron star can undergo
both electron and neutron scattering. There are, however, two distinct
classes of neutralinos, each of which is (mainly) produced in a
different proto-neutron star region. The two classes also have
different energy distributions, which is important, since the
mean-free path depends via the cross section on the neutralino
energy. Strictly speaking, we should average over the entire spectrum
of produced neutralino energies, but here we estimate the effect of
scattering by fixing the neutralino energy at the peak of the energy
distribution and evaluating cross sections, and hence mean-free
paths, at that value.

Thus for both electron and neutron scattering we apply the test
(\ref{eq:optdepth}) separately for two classes of neutralino:
\begin{enumerate}
\item Neutralinos produced by $e^+ e^- \rightarrow \lsp \lsp$. In
Sec.~\ref{ep}, we found that most of these are born in the outermost
10\% of the supernova. Thus in the case of these ``annihilation
neutralinos'' we take $r_0=0.9\, R_c$ in Eq.~(\ref{eq:optdepth}).  As
for their effective energy, in $e^+e^-\ra \lsp\lsp$, we must average
over the thermal {\it electron} energies to obtain the neutralino
energy. We estimate this as follows.  The neutralino production cross
section is proportional to $E^2_\lsp$.  Combining this with the energy
dependence of phase space ($E^2_\lsp$) and the Boltzmann suppression
$\exp(-E_\lsp/T_c)$ we expect a maximum number of neutralino pairs
with a {\it combined} energy $\langle E_{\lsp \lsp}^{\rm tot}
\rangle= 4\,T_c$. Thus for the average {\it individual} annihilation
neutralino energy we adopt $\langle E_\lsp\rangle\approx 2 T_c$. (For
$M_\lsp >2T_c$ we just set $E_\lsp=M_\lsp$.)

\item Neutralinos produced by binostrahlung. In this case the
neutralinos are predominantly produced in the centre of the supernova
and so we apply Eq.~(\ref{eq:optdepth}) with $r_0=0$.  By a similar
argument to that of the previous paragraph, we estimate $\langle
E_{\lsp\lsp}^{\rm tot}\rangle = 5T_c$, and thus for the individual
neutralinos $\langle E_\lsp\rangle = \frac{5}{2}T_c$.
\end{enumerate}

In both cases the produced neutralinos can scatter off either
electrons or nucleons. Thus, for each type of neutralino we now
determine separate bounds from $e \lsp$ and $n \lsp$ scattering on
$M_{\tilde e}$ and $M_{\tilde q}$ respectively.

\subsubsection{Neutralino-Electron Scattering}

The scattering cross section for a pure bino on electrons is given
by\footnote{We obtain the scattering for a massless photino by setting
  $M_\lsp=0$ and multiplying by $2\cos^4\theta_w/(Y(e_L)^4+Y(e_R)^4)$,
  where $Y(e_L)=-1/2$ and $Y(e_R)=-1$. The result agrees with
  Ref.~\cite{Lau:vf}.}:
\begin{equation}
\sigma(e+{\lsp}\ra e+{\lsp})=
\frac{17\pi\alpha^2}{12\cos^4\theta_w}\frac{s}{ M_{{\tilde e}}^4}
 \left(1-\frac{M_\lsp^2}{s}\right)^2\left(1+\frac{M_\lsp^2}{2s}
\right)^2\,,
\label{eq:echiscatt}
\end{equation}
where $s=M_\lsp^2+2E_e(E_\lsp-|\vec{p}_\lsp|\cos\theta)$, is the
centre-of-mass energy squared. The neutralino mean-free path
$\lam_\lsp$ depends strongly on the selectron mass, $M_{\tilde e}$,
and, less strongly, on $M_\lsp$.  For a small selectron mass, we obtain a
large cross section and thus a small mean-free path. In the following,
for a given neutralino mass, we shall determine a lower bound on the
selectron mass above which free-streaming occurs.

Estimating the electron energy as $E_e=\frac{3}{2}T$, we find that for
free-streaming of annihilation neutralinos to occur, the selectron
mass must obey:

\vspace{0.5cm}

\centerline{
\begin{tabular}{|c|c|c|c|c|c|c|c|}\hline 
$M_\lsp({\rm MeV})$ \phhigh  &        0  &10 &30  &50  &100 &150 &200 \\ \hline 
$M_{\tilde e} ({\rm GeV})>$\phhigh & 322 &322&321 &315 &341 &363 &377 \\
\hline
\end{tabular}}
\vspace{0.5cm}
\noindent where we have written the results for values of $M_\lsp$ between
0 and 200 MeV. The fact that the bound first drops and then rises with
rising $M_\lsp$ is due to the functional dependence of the cross
section (\ref{eq:echiscatt}).

$M_{\tilde e}$ must be larger than about $320-380\gev$ in order to
ensure free-streaming. Thus even though Sec.~\ref{ep} suggested that
selectron masses of order 200 GeV were forbidden by the SN1987A data
(provided that $M_\lsp < 200\mev$) this is not strictly speaking the
case, since if the selectron mass were that low then the neutralinos
produced by electron-positron annihilation would propagate diffusively
out of the supernova. To get an accurate bound for $M_{\tilde e} \sim
200$ GeV, neutralinos would have to be coupled to matter and included
in a supernova simulation.

There is thus a window of allowed selectron masses $100\gev
\lsim M_{\tilde e} \lsim 300\gev$ where interactions with electrons
mean that the annihilation neutralinos will diffuse out of the supernova.
This physics is beyond the scope of this paper and we have nothing to
add to the previous approximate computations of this regime performed
in Refs.~\cite{Grifols:fw,Ellis:1988aa}.

Meanwhile, a neutralino produced via binostrahlung can also scatter
off the supernova electrons. These neutralinos free stream as long as:

\vspace{0.5cm}

\centerline{
\begin{tabular}{|c|c|c|c|c|c|c|c|}\hline 
$M_\lsp({\rm MeV})$ \phhigh  & 0&10&30&50&100&150& 200 \\ \hline 
$M_{\tilde e} ({\rm GeV})>$\phhigh & 606 & 606 & 604 & 596 & 607 &646 & 670 \\
\hline
\end{tabular}}

\vspace{0.5cm}

However, if $M_{\tilde e}$ is lower than 600 GeV copious numbers of
neutralinos will be produced by annihilation, and the diffusion of
binostrahlung neutralinos can be safely neglected. So the possibility
of an $M_{\tilde e}$ low enough to induce binostrahlung neutralino
diffusion is already excluded by our other arguments, unless $M_\lsp
\gsim 150\mev$. The bounds of Sec.~\ref{ep} do not preclude $M_{\tilde e}
\sim 650\gev$ for this heavy a neutralino, but if $M_\lsp \gsim 150$
MeV we will see below that binostrahlung cannot be used to set a bound
on the squark mass anyway. Thus the constraint on $M_{\tilde e}$ from
the consideration of binostrahlung neutralinos is essentially
irrelevant.

\subsubsection{Neutralino-Neutron Scattering}
The neutralinos can also be trapped via neutralino-nucleon scattering,
Eq.~(\ref{scatter2}).  In order to compute the cross section, we can
employ the effective Lagrangian Eq.~(\ref{eff-lag-nuc}). We obtain
\begin{equation}
\sigma(\lsp+n\ra\lsp+n)=\frac{G_{SUSY}}{16\pi} \left[
{c_{\rm v}^\chi}^2 (E_\lsp^2-M_\lsp^2)+ 3{c_{\rm a}^\chi}^2
(E_\lsp^2+M_\lsp^2)\right]\,.
\label{xsec-nchi}
\end{equation}
We have made the approximation for the centre-of-mass energy squared
$s\approx M_n^2$, where $M_n$ is the neutron mass, and assumed
non-relativistic neutron spinors.  We see that we obtain contributions
from both the vector and axial-vector neutralino current. The
vector-current contribution vanishes in the limit of non-relativistic
neutralinos---a limit that is taken, for example, in computations
relevant to dark-matter detection.

For the case of neutralinos produced via neutralinostrahlung, we take
$\langle E_\lsp\rangle=\frac{5}{2}T_c$. Setting $r_0=0$ and using the
results from the LO-QPM, we then obtain the lower limit on the squark
mass

\vspace{0.5cm}

\centerline{
\begin{tabular}{|c|c|c|c|c|c|c|c|}\hline 
$M_\lsp({\rm MeV})$ \phhigh  & 0&10&30&50&100&150&200 \\ \hline 
$M_{\tilde q} ({\rm GeV})>$\phhigh & 306 &305& 298 &280 &269&330&381 \\
\hline
\end{tabular}}

\vspace{0.5cm}

Comparing this with the results in Sec.~\ref{nn}, we see that even
for the case of most dramatic neutralino-cooling
effects---$M_\lsp=0$---we can only exclude a region $300 \gev \leq
M_{\tilde q} \leq 350 \gev$ by the arguments presented there.  For
$M_\lsp \geq 30 \mev$, the values of $M_{\tilde q}$ excluded by the
supernova-cooling analysis of Sec.~\ref{nn} are so low that the
supernova neutralinos would be trapped by their strong interactions
with neutrons in the proto-neutron star. 

It is interesting to note that scattering is competitive with
production---even for $M_\lsp=0$---partly because of the sizeable
ratio of vector to axial-vector coupling constants.  The large value
of $c_{\rm v}^\chi/c_{\rm a}^\chi$ also means that if $M_\lsp \lsim
\frac{5}{2} T_c$ the cross section (\ref{xsec-nchi}) is dominated by
the term proportional to ${c_{\rm v}^\chi}^2$. This piece drops with
$M_\lsp$, but once $M_\lsp \gsim \frac{5}{2} T_c$ the cross section
begins to rise with $M_\lsp$ as the axial contribution takes
over. This interplay of vector and axial-vector pieces produces the
dependence of the $M_{\tilde q}$-bound on $M_\lsp$ seen in the two
tables of this subsection.

Annihilation neutralinos can also scatter off nucleons. For these we
again take $\langle E_\lsp \rangle = 2\,T_c$ and
$r_0=\frac{9}{10}R_c$, thereby obtaining the free-streaming
conditions:

\vspace{0.5cm}

\centerline{
\begin{tabular}{|c|c|c|c|c|c|c|c|}\hline 
$M_\lsp({\rm MeV})$ \phhigh  & 0&10&30&50&100&150&200 \\ \hline 
$M_{\tilde q} ({\rm GeV})>$\phhigh & 154 &153& 147 &132 & 152&186 &214 \\
\hline
\end{tabular}}

\vspace{0.5cm}

Thus, if $M_{\tilde q}$ is sufficiently small the arguments of
Sec.~\ref{ep} cannot be used to set bounds on $M_{\tilde e}$, since,
below the values of $M_{\tilde q}$ listed in this table, annihilation
neutralinos will propagate diffusively out of the core because of
strong neutron-neutralino scattering.

\subsection{Gravitational Trapping}
Interactions with matter are not the only way that neutralinos can
be trapped in the supernova. Neutralinos produced in the outer
regions whose kinetic energy obeys
\begin{equation}
E_{\rm kin} \leq \frac{G M_R M_\lsp}{R},
\end{equation}
will be trapped by gravitational attraction. $G$ is Newton's constant
and $R$ is the radius at which the neutralino is produced. $M_R$ is
the enclosed mass of the supernova at radius $R$. To get an estimate of
the effect of gravitational trapping we take $R \sim R_c$ and $M_R
\sim M_{\rm SN}$. In terms of neutralino momenta, this means that
neutralinos with $|{\vec {\rm \bf p}}_\lsp|$ small enough that:
\begin{equation}
\gamma=\sqrt{1 + \frac{|{\vec {\rm \bf p}
}_\lsp|^2}{M_\lsp^2 }} \leq 1 + \frac{GM_{\rm SN}}{R_c}
\label{eq:relcond}
\end{equation}
will not escape from the supernova. Here $\gamma=1/\sqrt {1-v^2}
$. Since $GM_{SN}/R_c\approx 0.15$, this demonstrates that
relativistic neutralinos (almost) always escape. 

In the {\it non}-relativistic case, it is easy to rewrite
Eq.~(\ref{eq:relcond}) as a condition on the neutralino velocity:
\begin{equation}
v^2 \leq \frac{R_S}{R_c},
\label{non-relcond}
\end{equation}
with $R_S=2 G M_{SN}\approx4.1\,{\rm km}$ the Schwarzschild radius of
the core. Assuming that the distribution of $\lsp$ velocities is
Maxwellian ($\langle E_{\rm kin}\rangle = 3T_c/2$), we find for
the non-relativistic neutralinos:
\begin{equation}
\langle v^2 \rangle =\frac{3T_c}{M_\lsp}\,.
\end{equation}
Combining this with Eq.~(\ref{non-relcond}) we obtain a lower bound
for the mass of gravitationally-bound neutralinos:
\begin{equation}
M_\lsp \geq \left(\frac{R_c}{R_S}\right) \, 3\, T_c\approx285\mev\,.
\end{equation}
This is outside the range of neutralino masses where we can set bounds
($M_\lsp\lsim 200\mev$) and thus our previous analysis holds.

\section{Conclusions}
\label{conc}

In general in a supernova, neutralinos will be produced by both
$e^+e^-$ annihilation and $NN$ binostrahlung. If the neutralinos
free-stream out of the supernova, they can lead to excessive cooling
which would alter the observed neurtino signal from SN1987A. In order
to exclude this, we set bounds on the relevant supersymmetric masses:
$(M_\lsp,\,M_{\tilde e})$ and $(M_\lsp,\,M_{ \tilde q})$,
respectively. The bounds are significantly stricter for
$e^+e^-$ annihilation, since the production cross
section is larger in that case.

For $e^+e^-$ annihilation we can, for a given neutralino mass, exclude
the following values of selectron mass, $M_{\tilde e}$:

\vspace{0.5cm}

\centerline{
\begin{tabular}{|c|c|c|c|c|c|c|c|}\hline 
$M_\lsp({\rm MeV})$ \phhigh  &        0  &10   &30   &50   &100 &150 &200 \\ \hline 
$M_{\tilde e} ({\rm GeV})<$\phhigh & 1275&1275 &1260 &1188 &930 &700 &450  \\
$M_{\tilde e} ({\rm GeV})>$\phhigh & 322 &322  &321  &315  &341 &363 &377 \\
\hline
\end{tabular}}

\vspace{0.5cm}

\noindent
If the selectron mass is below the lower bound, the neutralinos
produced by annihilation do not free stream, and so the arguments of
this paper do not apply. Meanwhile, if $M_{\tilde e}$ is above the
upper bound then an insuffcient number of neutralinos are produced to
affect the neutrino signal from SN1987A.

For the case of the squarks we obtain a similar table

\vspace{0.5cm}

\centerline{
\begin{tabular}{|c|c|c|c|}\hline 
$M_\lsp({\rm MeV})$ \phhigh  &        0  &10   &30    \\ \hline 
$M_{\tilde q} ({\rm GeV})<$\phhigh & 360 &351 &299  \\
$M_{\tilde q} ({\rm GeV})>$\phhigh & 306 &305  &298   \\
\hline
\end{tabular}}

\vspace{0.5cm}

\noindent
Only a small interval of $M_{\tilde q}$ values is excluded, even
for $M_\lsp=0$, because there is only a narrow region where
the coupling of the neutralinos to neutrons is large enough to
compete with neutrino production, but small enough to avoid diffusive
neutralino propagation.

Furthermore, we expect that many-body effects will modify this
squark-mass bound. Recent calculations of these effects in
the case of neutrinostrahlung~\cite{vD03} suggest that the dominant
many-body correction to the binostrahlung rate at supernova
temperatures will be the Landau-Pomeranchuk-Migdal (LPM) effect. The
LPM effect will tend to suppress the production of axial radiation,
thereby serving to still further weaken our already weak squark-mass
bound~\cite{HPR,vD03,Ke97,RS95}.

\begin{figure}[h!]
\vspace*{10cm}
\includegraphics{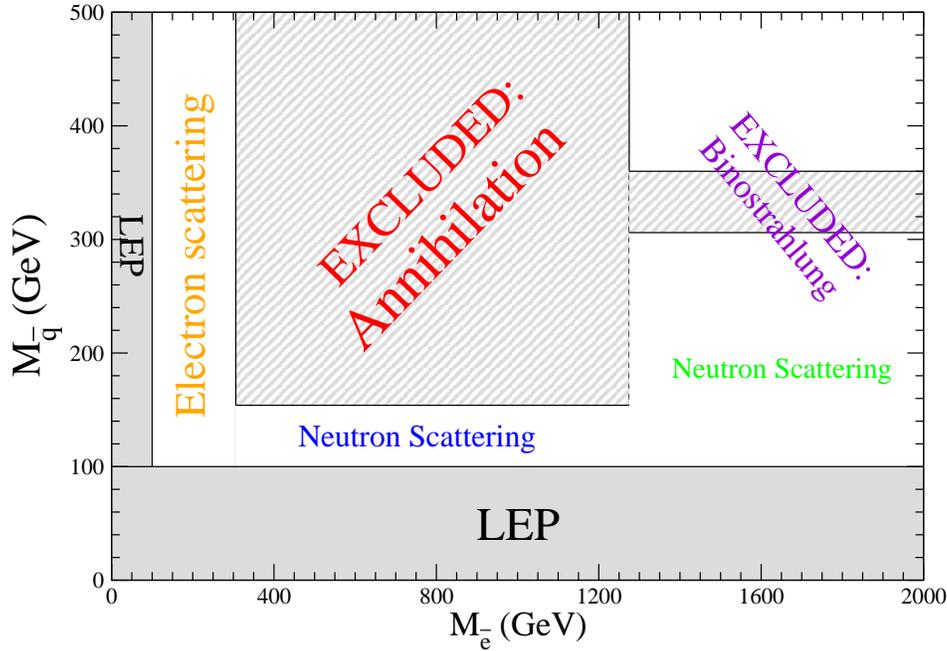}
\caption{{\it $M_{\tilde e}-M_{\tilde q}$ parameter space, with the
regions excluded by the arguments of this paper for the case
$M_\lsp=0$ shaded. The regions already explored by LEP searches are
shown in grey. The white regions indicate areas of parameter space in
which we are unable to set bounds. When this is because of diffusive
propagation of the neutralino we have indicated the mechanism that
leads to diffusion. Values of $(M_{\tilde e},M_{\tilde q})$ in the
upper-right-hand corner are permitted because there the neutralino
production rate is too low to affect the SN1987A neutrino signal
significantly. }}
\label{combi}
\end{figure}

In Fig.~\ref{combi} we present the combined bounds for $M_\lsp=0$.
Supernova-cooling arguments exclude a large range of possible
selectron masses if the neutralino is massless, but $M_{\tilde e}
\gsim 1300\gev$ is still allowed.  In the region $M_{\tilde e} \lsim
300 \gev$ annihilation neutralinos diffuse due to electron
scattering. They also diffuse, but this time due to interactions with
neutrons, if $M_{\tilde q} \lsim 150 \gev$.  We cannot set bounds in
these regions of parameter space. Their proper treatment would require
a complete proto-neutro-star simulation.

\vspace{0.5cm}

The bounds shown in Fig.~\ref{combi} do not change significantly with
$M_\lsp$ until masses of 50 MeV are reached. Above this value there is
no squark mass bound whatsoever from SN1987A. Selectron mass bounds
are also significantly reduced. For neutralinos with mass $M_\lsp\gsim
200\mev$, we can essentially place no limits. Once this range of
$M_\lsp$ values is reached the LEP and Tevatron bounds on the sfermion
masses, together with the Boltzmann suppression of heavy-particle production,
guarantee that the supernova neutralino production rate is just too
small to modify the neutrino signal markedly.

In summary, we showed that a light neutralino, $M_\lsp<{\cal
O}(1\gev)$ is allowed by laboratory experiments and consistent with
the MSSM if it is pure bino. We have investigated the bounds obtained
from the SN1987A neutrino signal on a quasi-stable neutralino with
mass $M_\lsp<200\mev$ and studied the production of such neutralinos
in both electron-positron annihilation and nucleon-nucleon
collisions. In the former case, we compared the Raffelt criterion with
an estimate of the integrated emitted energy calculated using the
radial temperature and degeneracy profiles obtained in supernova
simulations. We found good agreement between the two approaches. We
then employed the Raffelt criterion to bound neutralino production in
nucleon-nucleon collisions inside the supernova.  Overall we find that
for selectron masses $300 \gev \lsim M_{\tilde e} \lsim 900 \gev$ we
can exclude neutralino masses below $100\mev$. On the other hand, for
selectron masses above $1200\gev$ there is no lower bound on the
lightest neutralino mass. There is also a near absence of any bound on
the squark masses: only a narrow range of $M_{\tilde q}$ can be
excluded, and this only for $M_\lsp \lsim 20 \mev$.
 
\section*{Acknowledgments}
H.~D. would like to thank Subir Sarkar and Graham Ross for initial
discussions leading to this work.  C.~H. and D.~R.~P. would like to
thank Sanjay Reddy for the previous collaboration on which part of
this work is based. We all thank Sanjay Reddy for detailed discussions
on the supernova aspects of this paper.  We thank W. Vogelsang for
discussions on polarized structure functions and Manuel Drees and
Ricardo Flores for discussions on the effective quark-neutralino
Lagrangian. D.~R.~P. thanks the Forschungszentrum, J\"ulich for
hospitality during the early stages of this work. The work of
D.~R.~P. was supported by the U.~S. Department of Energy under grants
DE-FG02-93ER40756 and DE-FG02-02ER41218.

\appendix

\section{Appendix}

In this appendix we give the explicit result for the angle-averaged
hadronic tensor, $\bar{\cal H}_{ii}$, in terms of phase shifts.  The
quantity needed for the leading order diagrams is the nucleon--nucleon
scattering T--matrix $T_{NN}$ in plane wave decomposition. It can be
decomposed into partial waves in the following way. (We only give the
expressions for equal final and initial spin.  This is all that is
required for the reaction $nn \to nn\lsp\lsp$.)
\begin{equation}
\langle S \, M_S', {\bf p} \, '|T_{NN}|S \, M_S, {\bf p} \rangle =
\sum Y_{L'M_L'}({\bf p} \, ')Y_{LM_L}({\bf p})^*
{\cal T}(p,S' \, M_S', \, S \, M_S, \, L', \, \Delta L, M_J) \ ,
\end{equation}
where
\begin{equation}
{\cal T}(p,S' \, M_S', \, S \, M_S, \, L', \, \Delta L, M_J)
 = \sum \langle S' \, M_S', \, L' \, M_L'|J \, M_J \rangle
\langle S \, M_S, \, L \, M_L|J \, M_J \rangle \
T^{JL'LS}(p) \ .
\end{equation}
The quantity $T^{JL'LS}(p)$ can be related to the phase shifts deduced
from nucleon--nucleon scattering data directly. For example, in the
case of non-coupled channels we have
\begin{equation}
T^{\ell}(p)=(2\pi)^6\left(\frac{2}{\pi M p}\right)
e^{i\delta_\ell (p)}\sin (\delta_\ell (p)) \ ,
\end{equation}
where $\delta_\ell$ denotes the phase shift in the partial wave
denoted by $\ell$. These quantities are available online \cite{VPI}. 
Note, only elastic scattering is kinematically allowed under the
conditions of the study presented.

It is straight forward yet lengthy to relate $\cal T$ to the angle-averaged
hadronic tensor $\bar{\cal H}_{ii}$ as defined in Eq.~(\ref{hbardef}):
\begin{eqnarray}
\nonumber
\bar{\cal H}_{ii} &=& \sum 2 \left\lbrace
 (2-M_SM_S')|{\cal T}(p,1 \, M_S', \, 1 \, M_S, \, L', \, \Delta L, M_J)|^2
\right. \\
& & \ \qquad
\left.-2\mbox{Re}\left[{\cal T}(p,1 \, (M_S'-1), \, 1 \, M_S, \, L', \,
\Delta L, M_J-1)^* \right.\right.\\ && \left.\left. \qquad\qquad\qquad
\times{\cal T}(p,1 \, M_S', \, 1 \, (M_S+1), \, L', \, \Delta L, M_J)\right]
\right\rbrace \ . \nonumber
\label{hbarres}
\end{eqnarray}


\end{document}